\documentclass[twocolumn]{aastex63}

\usepackage{xcolor,soul}
\sethlcolor{green}
\received{June 1, 2019}
\revised{January 10, 2019}
\accepted{\today}
\submitjournal{ApJ}

\shorttitle{Formation location of HD 209458b}
\shortauthors{Dash et al.}
\graphicspath{{./}{figures/}}

\begin{document}

\title{Linking atmospheric chemistry of the hot Jupiter HD 209458b to its formation location through infrared transmission and emission spectra}

\correspondingauthor{Liton Majumdar, Spandan Dash}
\email{dr.liton.majumdar@gmail.com, dashspandan@gmail.com}

\author{Spandan Dash}
\affiliation{Department of Physics, University of Warwick, Coventry, West Midlands, United Kingdom, CV4 7AL}
\affiliation{School of Earth and Planetary Sciences, National Institute of Science Education and Research, Jatni 752050, Odisha, India}
\affiliation{Homi Bhabha National Institute, Training School Complex, Anushaktinagar, Mumbai 400094, India}

\author{Liton Majumdar}
\affiliation{School of Earth and Planetary Sciences, National Institute of Science Education and Research, Jatni 752050, Odisha, India}
\affiliation{Homi Bhabha National Institute, Training School Complex, Anushaktinagar, Mumbai 400094, India}


\author{Karen Willacy}
\affiliation{Jet Propulsion Laboratory, California Institute of Technology, 4800 Oak Grove Drive, Pasadena, CA 91109, USA}

\author{Shang-Min Tsai}
\affiliation{Atmospheric, Ocean, and Planetary Physics, Department of Physics, University of Oxford, Oxford, OX1 3PU, United Kingdom}

\author{Neal Turner}
\affiliation{Jet Propulsion Laboratory, California Institute of Technology, 4800 Oak Grove Drive, Pasadena, CA 91109, USA}


\author{P. B. Rimmer}
\affiliation{MRC Laboratory of Molecular Biology, Francis Crick Ave, Cambridge CB2 0QH, United Kingdom}
\affiliation{Astrophysics Group Cavendish Laboratory, JJ Thomson Ave, Cambridge CB3 0HE, United Kingdom}
\affiliation{Department of Earth Sciences, University of Cambridge, Downing St, Cambridge CB2 3EQ, United Kingdom}


\author{Murthy S. Gudipati}
\affiliation{Jet Propulsion Laboratory, California Institute of Technology, 4800 Oak Grove Drive, Pasadena, CA 91109, USA}

\author{Wladimir Lyra}
\affiliation{Department of Astronomy, New Mexico State University, PO BOX 30001, MSC 4500, Las Cruces, NM 88003-8001}

\author{Anil Bhardwaj}
\affiliation{Physical Research Laboratory, Ahmedabad, 380009, India}




\begin{abstract}

The elemental ratios of carbon, nitrogen, and oxygen in the atmospheres of hot Jupiters may hold clues to their formation locations in the protostellar disc. In this work, we adopt gas phase chemical abundances of C, N and O from several locations in a disc chemical kinetics model as sources for the envelope of the hot Jupiter HD 209458b and evolve the planet’s atmospheric composition using a 1D chemical kinetics model, treating both vertical mixing and photochemistry. We consider two atmospheric pressure-temperature profiles, one with and one without a thermal inversion. From each of the resulting 32 atmospheric composition profiles, we find that the molecules CH$_4$, NH$_3$, HCN, and C$_2$H$_2$ are more prominent in the atmospheres computed using a realistic non-inverted P-T profile in comparison to a prior equilibrium chemistry based work which used an analytical P-T profile. We also compute the synthetic transmission and emission spectra for these atmospheres and find  that many spectral features vary with the location in the disc where the planet's  envelope was accreted. By comparing with the species detected using the latest high-resolution ground-based observations, our model suggests HD 209458b could have accreted most of its gas between the CO$_2$ and CH$_4$ icelines with a super solar C/O ratio from its protostellar disc, which in turn directly inherited its chemical abundances from the protostellar cloud. 
Finally, we simulate observing the planet with the James Webb Space Telescope (JWST) and show that differences in spectral signatures of key species can be recognized. Our study demonstrates the enormous importance of JWST in providing new insights into hot Jupiters’ formation environments.
\end{abstract}



\section{Introduction}
The detection of the Na doublet at 589.3 nm in the atmosphere of HD 209458b \citep{charbonneau2002detection} heralded the era of atmospheric characterization of exoplanets and confirmed theoretical predictions made by \citet{seager2000theoretical, brown2001transmission, hubbard2001theory}. This was followed by the  discovery of an extended atmosphere on the same planet with the detection of HI, OI and CII in the escaping upper atmosphere beyond the Roche lobe \citep{vidal2003extended,vidal2004detection}. The source of HI was almost immediately explained by a simple hydrogen/oxygen based photochemical model by \citet{liang2003source}. Since then more observations have resulted in the detection of simple molecules like H$_2$O, CO, CH$_{4}$, HCN and NH$_3$ in HD 209458b's atmosphere \citep{swain09, sing2016continuum, tsiaras2016new, tsiaras2018population, brogi2019retrieving, snellen2010orbital, macdonald2017hd, sanchez2019water, Giacobbe2021}. Chemical models to explain spectral observations have kept pace with both chemical equilibrium e.g. TEA \citep{blecic2016tea} and FastChem \citep{stock2018fastchem} as well as chemical kinetics models e.g. \citet{zahnle2009atmospheric}, KINETICS \citep{moses11, moses13a, moses13b}, Kasting Model \citep{kopparapu12}, Venot Model \citep{venot2012chemical}, ATMO \citep{amundsen2014accuracy, drummond2016effects}, VULCAN \citep{tsai2017vulcan}, ARGO \citep{rimmer2016chemical}, LEVI \citep{hobbs2019chemical} utilizing better and more precise networks of chemical reactions aided by new and/or improved laboratory data as well as observations. 
\\
\\
Formation conditions and evolution history of exoplanets can be potentially linked to their C/O and N/O ratios \citep{oberg2011effects, mordasini2016imprint, cridland2019connectinga, eistrup2016setting, eistrup2018molecular, notsu2020composition, turrini2021tracing, hobbs2021molecular, ali2017disentangling, nowak2020peering}. For giant planets formed by the \emph{Core Accretion} mechanism, the core is thought to be formed by accretion of solid ice material from the disc midplane by a core accretion or pebble accretion stage or a hybrid mix of both \citep{alibert2018formation}, and the envelope is thought to be formed by rapid/runaway accretion of gas from the local gas disc region close to the disc midplane after the formation of their cores \citep{eistrup2016setting, notsu2020composition}. Envelope accretion uses gas available vertically for at least 1 scale height \citep{tanigawa2012distribution, morbidelli2014meridional, teague2019meridional} from the disc midplane and \citet{cridland2020impact} found that planets formed within 20 au would reflect the C/O ratio of the gas accreted during this stage. The rapid gas accretion timescale is also much smaller than Type-II migration timescale required for the cores to migrate inwards \citep{pollack1996formation} and hence almost all the gas in the planet's envelope is accreted very close to where its runaway accretion starts.
\\
\\
Local gas phase and grain abundances of elements like C, O and N differ in the radial direction as shown by disc mid plane chemistry models \citep{oberg2011effects, eistrup2016setting, eistrup2018molecular}. In a simplified scenario, these radially varying profiles of elemental abundances will then serve as the initial conditions for further evolution of the core as well as the envelope. Hence, looking at the evolution of such a gaseous envelope has the potential to link an exoplanet at the very least to the location where it accreted the majority of its gaseous envelope.  However, in a more realistic sense, the envelope itself can be contaminated by accretion of more icy planetesimals or comets and asteroids or even be affected by degassing from the differentiating and partitioning core. All these processes can then muddle the entire scenario such that it becomes very difficult to determine a direct connection of chemistry of gas giant atmospheres with the disc chemistry at the location where they formed. In this work, we assume the earlier simpler case but we discuss the effect of disc solid contamination as a limitation of the model (see Section 4.3) and consider one possible model of contamination from \citet{turrini2021tracing}'s work and find the effect it can have on spectra. The simpler case has also been assumed by previous works which include disc chemistry in planet formation and atmospheric evolution models \citep{cridland2017composition, cridland2019connectinga, cridland2019connectingb, cridland2020impact, booth2019planet}. However, all their models had a limited set of chemical networks or used networks of an equivalent complexity as \citet{eistrup2016setting}. This motivates us to use the results from \citet{eistrup2016setting} as the starting point for our analysis.
\\
\\
\citet{mordasini2016imprint} modeled the effect of disc chemistry on simulated JWST and ARIEL spectral signatures of hot Jupiters using the concept outlined above in a \emph{planetesimal accretion regime} with the final atmospheric composition being enriched by planetesimal impacts during formation. They found that their hot Jupiters could be divided into `dry' or `wet' depending on formation inside or outside the water iceline respectively. Both types were depleted in C and enriched in O as the standard inner disc chemistry model assumed by them was already depleted in C. This resulted in `dry' planets with C/O ratios less than 0.2 and `wet' planets with C/O ratios 0.1-0.5 respectively. They also found differences in their modelled JWST and ARIEL spectra for the extreme cases of `dry' and `wet' hot Jupiters. C/O ratios greater than 1 for planets were only possible for `dry' hot Jupiters if the disc chemistry was assumed to be non-standard with no refractory C depletion. However, without knowledge of how this could happen, it was difficult to find formation location of Hot Jupiters that have evidence of high (including super-solar) atmospheric C/O ratios e.g. HD 209458b \citep{Giacobbe2021}. Nonetheless, the basis of linking exoplanet composition through transmission and emission spectra to disc chemistry still remains important and forms a major basis of this work.
\\
\\
\citet{eistrup2016setting} presented the case for including detailed gas phase and grain networks to evolve the chemistry of gas and ice in disc midplanes and found that the results for the radially varying gaseous elemental abundance profiles of CO, CO$_2$, H$_2$O, NH$_3$, N$_2$, O$_2$ and CH$_4$ were more diverse. Using this as basis, \citet{notsu2020composition}, for the first time, presented cases of model Hot Jupiter atmospheric chemistry starting from gas phase elemental abundances obtained from such disc chemistry models (within the limit of 20 au provided in \citet{cridland2020impact}) and found cases with gas phase envelope C/O ratios greater than super-solar and $>$1 as well. They assumed that local gas accretion occurs for a migrating hot Jupiter in a narrow region in between various icelines (H$_2$O, CO$_2$, CH$_4$ and CO) and hence the elemental abundances of C/H, O/H, N/H and the overall C/O ratios would be an indicator of formation location of a Hot Jupiter. However, they did not look at how changes in initial elemental abundances based on disc location might produce detectable observational spectral signatures.
\\
\\
Following the potential evidence in \citet{mordasini2016imprint}'s work that JWST and ARIEL can be used to trace composition in exoplanet atmospheres, we extend \citet{notsu2020composition}'s work to do the same with one caveat: we simulate disequilibrium chemistry by including the effects of vertical mixing, diffusion and photochemistry instead of just equilibrium chemistry. \citet{moses2014chemical} noted that disequilibrium can greatly affect the levels of species such as HCN and C$_2$H$_2$ in the photosphere which can produce detectable spectral signatures. This limitation about usage of equilibrium over disequilibrium chemistry in their model was noted in \citet{notsu2020composition} itself. All these factors form the basis behind our study.
\\
\\
More recently, \citet{turrini2021tracing} and \citet{hobbs2021molecular} have also worked on tracing the history of planet formation through molecular markers. \citet{turrini2021tracing} have used initial gas and solids/ice abundances gleaned from a sophisticated analysis of meteoritic abundances, ISM abundances and astrochemical model abundances from \citet{eistrup2016setting} and \citet{eistrup2018molecular}. However, they have only focused on the \emph{Inheritance} with low ionization case (see Section 2.3 for an overview of the cases). We have used the \emph{reset} (atomic) and \emph{inheritance} (molecular) inputs, with both high and low ionization rates in this work. \citet{hobbs2021molecular} used a large grid of initial C/H, O/H, N/H abundances and P-T-K$_{zz}$ profiles to evolve their exoplanetary atmosphere using the chemical kinetics code LEVI and then compared the range of resultant molecular abundance values of H$_2$O, CO, CH$_4$, CO$_2$, HCN and NH$_3$ to the possible initial abundance values predicted by different planet formation, enrichment and migration models. In comparison, our study is restricted to only one possible planet formation and migration pathway. However, their chemistry from \citet{woitke2009radiation} is based on chemical equilibrium in the disc midplane as opposed to the time-dependent chemical kinetics of \citet{eistrup2016setting}. \citet{hobbs2021molecular} have also used the initial abundance values used in \citet{turrini2021tracing}, but as we pointed out earlier, it's only for the case of inheritance with low ionization. However, neither of the two studies directly looked at the possibility of linking spectra to the chemistry in the protostellar disc.
\\
\\
In this paper, we consider the specific case of the exoplanet HD 209458b and consider two different P-T profiles (with and without a temperature inversion). We use a custom built wrapper that couples the 1D chemical kinetics code VULCAN \citep{tsai2017vulcan, tsai2021comparative} (in order to simulate disequilibrium chemistry by incorporating vertical transport, diffusion and photochemistry) with a radiative transfer package petitRADTRANS \citep{molliere2019petitradtrans} (to create synthetic transmission and emission spectra). We vary the initial elemental abundances based on the values provided in \citet{notsu2020composition} for gas accretion at specific radial positions in the disc and determine if the differences in the constructed transmission and emission spectra could be an indicator of its C/O ratio and/or where the exoplanet could have accreted most of its gas from the disc.

\section{Methods}
\subsection{Physical structure of HD 209458b}
\begin{figure}
    \centering
    \includegraphics[width = \columnwidth]{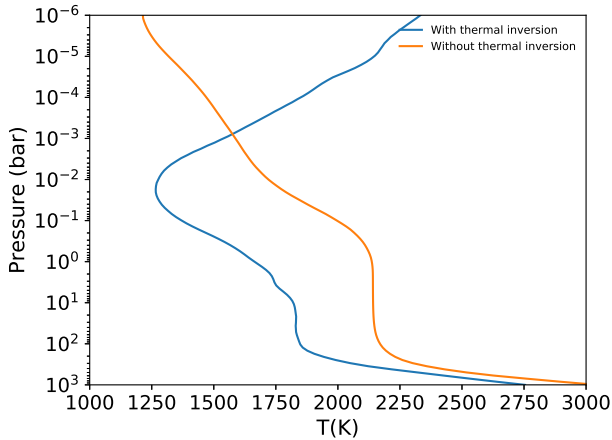}
    \caption{Both P-T profiles used in this study. For the case of thermal inversion, the P-T profile is as used in \citet{moses2011disequilibrium} and for the non inverted case, the temperature profile is calculated using the open source self-consistent radiative transfer code HELIOS using the planetary parameters for HD 209458b and solar spectrum as an analogue.}
    \label{fig:9}
\end{figure}

HD 209458b was the first exoplanet to be discovered using the transit method. It orbits around a 1.203 R$_{\odot}$ \citep{boyajian2015stellar} and 1.148 M$_{\odot}$ \citep{southworth2010homogeneous} star situated 47 pc away. Since these stellar parameter values are very close to the values used by \citet{notsu2020composition}, we chose HD 209458 as a viable candidate for this study. Additionally, HD 209458b's mass is 0.69 M$_{J}$, radius is 1.38 R$_{J}$ and orbital radius is 0.04747 au \citep{southworth2010homogeneous}. 
\\
\\
The P-T profile of this planet has been subject to debate with \citet{burrows2007theoretical} and \citet{knutson20083} initially
suggesting the existence of a thermal inversion in the atmosphere. However, more recently \citet{diamond2014new} and
\citet{line2016no} have obtained evidence against a thermal inversion being present between 1 bar to 10$^{-3}$ bar. Given this
uncertainty we use two P-T profiles, one with and one without a thermal inversion. Both P-T profiles are shown in Figure \ref{fig:9}.
\\
\\
The profile with a thermal inversion is taken from Figure 2 of \citet{moses2011disequilibrium}, and for this model we also use the Pressure-$K_{zz}$ profile from the same paper (see Figure 1 of \citet{moses2011disequilibrium}). The profile without the inversion is constructed using the open source self-consistent radiative transfer code for exoplanet atmospheres called HELIOS \footnote{https://github.com/exoclime/HELIOS} \citep{malik2017helios} by considering H$_2$O, CH$_4$, CO, CO$_2$, NH$_3$, HCN, C$_2$H$_2$, NO, SiH, CaH, MgH, NaH, AlH, CrH, AlO, SiO, CaO, Na, K and H- as sources of opacities in the atmosphere. In addition, CIA opacities for H$_2$-H$_2$ and H$_2$-He were also used. This profile is different from the non-inverted P-T profile adopted by \citet{notsu2020composition} who calculated it using the analytical P-T formula from \citet{guillot2010radiative}. Notably, our P-T profile is cooler at a pressure $>$ 10$^2$ bar, hotter over the range of 10$^2$-10$^{-4}$ bar and cooler at pressures $<$ 10$^{-4}$ bar. Our non-inverted P-T profile is slightly hotter than the one presented in \citet{Lavvas21} since we have used a heat re-distribution factor (f) of  2/3 to represent the dayside in a poor heat-redistribution case \citep{Komacek16}. 
\\
\\
For generating synthetic transmission spectra, a reference photospheric pressure is also needed. We use a nominal value of atmospheric pressure at the photosphere (P$_{ref}$) = 10$^{-2}$ bar or 10 mbar. This is about 3 times higher than the value obtained by  \citet{macdonald2017hd} who found the photosphere to lie at a mean value of 10$^{-2.45}$ bar using retrieval from transmission spectrum observations.

\subsection{Chemical evolution and spectra}
We use a custom built Python3 wrapper that couples a chemical kinetics code (VULCAN) with a radiative transfer code (petitRADTRANS) to obtain transmission and emission spectra for any exoplanet of interest. We use the open source 1D chemical kinetics model VULCAN\footnote{https://github.com/exoclime/VULCAN} as the first part of this package to evolve the atmospheric chemistry for HD 209458b. We use the default NCHO network provided in the original repository which tracks over 600 reactions linking 50 species \citep{zilinskas2020atmospheric}. We include eddy diffusion ($K_{zz}$), molecular diffusion and photochemistry while running all simulations, effectively utilizing disequilibrium chemistry for our exoplanet candidate. For the stellar flux, we use the default flux file for the Sun provided in the original repository (Gueymard solar flux file) as both our Sun and HD 209458 are G type stars with similar mass and radius parameters.
\\
\\
For the NCHO network, VULCAN uses specific elemental abundances for N, C, O and He with respect to H in order to first calculate the initial atmospheric molecular mixing ratios using the coupled equilibrium chemistry code FastChem\footnote{https://github.com/exoclime/FastChem} \citep{stock2018fastchem}. We retain the abundance of He with respect to H as 0.097 \citep{asplund2009chemical, tsai2017vulcan} and change the others to values which we discuss in Section 2.3 and are listed in Table \ref{table:1}.
\\
\\
We use the open source radiative transfer package petitRADTRANS\footnote{https://petitradtrans.readthedocs.io/en/latest/} \citep{molliere2019petitradtrans} to obtain synthetic transmission and emission spectra. Apart from using the stellar and planetary parameters as provided in Section 2.1, we use line opacities of H$_2$, CO, CO$_2$, CH$_4$, H$_2$O, HCN, NH$_3$, C$_2$H$_2$ and CN. In addition, we include CIA (Collision Induced Absorption) opacities for H$_2$-H$_2$ and H$_2$-He and Rayleigh scattering for H$_2$ and He. All of these are already present in the default latest version of petitRADTRANS. We run petitRADTRANS in the default \emph{c-k} (i.e. correlated-k) mode and hence our synthetic spectra have a resolution of R=1000. For the emission spectra, we use a synthetic PHOENIX stellar spectrum of $T_{eff}$ = 6092 K \citep{boyajian2015stellar} which is constructed using the default PHOENIX spectra calculator in petitRADTRANS.

\subsection{Elemental abundances}
\begin{deluxetable*}{ccccccccc}
\tablewidth{0pt}
\tablecaption{Input parameters used to run each atmospheric model with the first horizontal line separating molecular initial abundances above from atomic initial abundances below. Each model corresponds to an alphabet running from A to P and 1 or 2 represent use of a thermally inverted or a non thermally inverted P-T profile respectively. The models after the second horizontal line are initial abundances from Table 4 of \citet{turrini2021tracing} representing atmospheric enrichment by solids. \label{table:1}}
\tablehead{
\colhead{Model} & \colhead{Location (au)} & \colhead{Ionization} &  \colhead{C/H} & \colhead{O/H} & \colhead{N/H} & \colhead{C/O$^{a}$} & \colhead{C/N} & \colhead{Initial abundances}}
\startdata
A1/A2 & 0.5 & low & 1.81$\times$10$^{-4}$ & 5.20$\times$10$^{-4}$ & 6.24$\times$10$^{-5}$ & 0.35 & 2.90 & Molecular\\
B1/B2 & 1.0 & low & 1.67$\times$10$^{-4}$ & 2.07$\times$10$^{-4}$ & 6.24$\times$10$^{-5}$ & 0.81 & 2.68 & Molecular\\
C1/C2 & 5.0 & low & 7.06$\times$10$^{-5}$ & 5.57$\times$10$^{-5}$ & 4.18$\times$10$^{-5}$ & 1.27 & 1.69 & Molecular\\
D1/D2 & 20.0 & low & 5.40$\times$10$^{-5}$ & 5.37$\times$10$^{-5}$ & 4.14$\times$10$^{-5}$ & 1.00 & 1.30 & Molecular\\
E1/E2 & 0.5 & high & 1.80$\times$10$^{-4}$ & 5.19$\times$10$^{-4}$ & 6.24$\times$10$^{-5}$ & 0.35 & 2.90 & Molecular\\
F1/F2 & 1.0 & high & 1.81$\times$10$^{-4}$ & 2.20$\times$10$^{-4}$ & 6.24$\times$10$^{-5}$ & 0.82 & 2.90 & Molecular\\
G1/G2 & 5.0 & high & 1.76$\times$10$^{-5}$ & 1.87$\times$10$^{-5}$ & 4.68$\times$10$^{-5}$ & 0.94 & 0.38 & Molecular\\
H1/H2 & 20.0 & high & 3.07$\times$10$^{-6}$ & 3.07$\times$10$^{-6}$ & 3.99$\times$10$^{-5}$ & 1.00 & 0.08 & Molecular\\
\hline
I1/I2 & 0.5 & low & 1.81$\times$10$^{-4}$ & 5.21$\times$10$^{-4}$ & 6.24$\times$10$^{-5}$ & 0.35 & 2.90 & Atomic\\
J1/J2 & 1.0 & low & 1.81$\times$10$^{-4}$ & 5.17$\times$10$^{-4}$ & 6.24$\times$10$^{-5}$ & 0.35 & 2.90 & Atomic\\
K1/K2 & 5.0 & low & 3.16$\times$10$^{-5}$ & 2.37$\times$10$^{-4}$ & 5.65$\times$10$^{-5}$ & 0.13 & 0.56 & Atomic\\
L1/L2 & 20.0 & low & 6.19$\times$10$^{-6}$ & 1.51$\times$10$^{-5}$ & 2.43$\times$10$^{-5}$ & 0.41 & 0.25 & Atomic\\
M1/M2 & 0.5 & high & 1.81$\times$10$^{-4}$ & 5.21$\times$10$^{-4}$ & 6.24$\times$10$^{-5}$ & 0.35 & 2.90 & Atomic\\
N1/N2 & 1.0 & high & 1.81$\times$10$^{-4}$ & 5.19$\times$10$^{-4}$ & 6.24$\times$10$^{-5}$ & 0.35 & 2.90 & Atomic\\
O1/O2 & 5.0 & high & 8.16$\times$10$^{-5}$ & 3.25$\times$10$^{-4}$ & 5.82$\times$10$^{-5}$ & 0.25 & 1.40 & Atomic\\
P1/P2 & 20.0 & high & 3.90$\times$10$^{-6}$ & 4.36$\times$10$^{-6}$ & 2.34$\times$10$^{-5}$ & 0.89 & 0.17 & Atomic\\
\hline
T1 & 5 & low & 2.8$\times$10$^{-4}$ & 5.7$\times$10$^{-4}$ & 7.6$\times$10$^{-5}$ & 0.49 & 3.64 & Molecular\\
T2 & 12 & low & 4$\times$10$^{-4}$ & 6.9$\times$10$^{-4}$ & 8.1$\times$10$^{-5}$ & 0.58 & 4.95 & Molecular\\
T3 & 19 & low & 5.3$\times$10$^{-4}$ & 1$\times$10$^{-3}$ & 1$\times$10$^{-4}$ & 0.53 & 5.31 & Molecular\\
T4 & 50 & low & 9.7$\times$10$^{-4}$ & 2$\times$10$^{-3}$ & 1.4$\times$10$^{-4}$ & 0.50 & 6.80 & Molecular\\
T5 & 100 & low & 1.4$\times$10$^{-3}$ & 2.8$\times$10$^{-3}$ & 1.8$\times$10$^{-4}$ & 0.50 & 7.54 & Molecular\\
T6 & 130 & low & 2.2$\times$10$^{-3}$ & 4.5$\times$10$^{-3}$ & 2.7$\times$10$^{-4}$ & 0.50 & 8.28 & Molecular\\
\enddata
\tablenotetext{a}{Solar C/O is 0.55 \citep{Asplund09}. Accordingly, in this paper we have used sub-solar and super-solar nomenclature for C/O values less than and greater than this value.}
\end{deluxetable*}

We use the 16 values of elemental gas phase C/H, O/H and N/H abundances listed in Table 1 in \citet{notsu2020composition}. All these were derived from the midplane disc chemistry model of \citet{eistrup2016setting} which incorporates gas phase, gas-grain chemistry and grain surface chemistry and is evolved until 1 Myr to get these results. The iceline locations for H$_2$O, CO, CH$_4$ and CO$_2$ are assumed to lie at 0.7, 2.6, 16 and 26 au corresponding to temperatures of 177, 88, 28 and 21 K respectively. As with \citet{notsu2020composition}, we have also listed the initial elemental abundance values corresponding to different radial distances in the disc (0.5 au, 1 au, 5 au and 20 au corresponding positions interior to the H$_2$O iceline, between the H$_2$O and CO$_2$ icelines, between CO$_2$ and CH$_4$ icelines and between CH$_4$ and CO icelines respectively) in Table \ref{table:1}.
\\
\\
The first 8 values in Table \ref{table:1} are abundances of C, O and N with respect to H in a case where the disc wholly inherits the chemistry from the protostellar cloud (\emph{inheritance scenario} tagged as `molecular initial abundances') and the second part of the table is the case where the chemistry in the disc gas can be altered significantly from the initial protostellar cloud \citep{visser2015chemical} due to heating events generated from the protostar (e.g. stellar irradiation, accretion bursts, shocks generated during protostellar formation stage etc). \citet{eistrup2016setting} posited that an extreme case of such alteration could convert all molecules to atoms within 30 au which will then subsequently reform into molecules and solids in a condensation sequence. Hence, this case is labelled as \emph{reset scenario} or `atomic initial abundances'. Both situations represent two extreme positions for the initial chemical abundances of a disc formed from the collapse of a protostellar cloud. A realistic situation is hence expected to lie in between these two cases.
\\
\\
Each case is further subdivided into sub-cases where the disc is under low ($\zeta < 10^{-18}$) or high ($\zeta \sim 10^{-17}$) values of ionization corresponding to low and high level of chemical processing in the disc respectively. The low ionization condition takes into account ionization only from short lived decay products of radio nucleotides inside the disc itself. The high value of ionization takes into account both the previous source as well as cosmic ray ionization external to the disc. The first case produces a higher level of ionization in the inner disc as compared to the outer disc and the latter case has a higher level of ionization in the outer disc compared to the inner disc. Which case works in a disc ultimately depends on how much the disc is shielded by solar flares and magnetic fields from external cosmic rays \citep{eistrup2016setting}.
\\
\\
For the case of molecular initial abundances with low ionization in Table \ref{table:1}, the gas-phase C/O ratio increases as we move from the inner disc to the outer disc and cross the various icelines. This is very similar to the step-wise C/O distribution in \citet{oberg2011effects} and shows that this case has minimal effect of chemical processing in the disc. The case for molecular initial abundances with high ionization in the same table also shows the same behaviour except for a difference between the CO and CH$_4$ icelines at 5 au where the C/O ratio is now sub-solar (0.94 vs 1.27 for the previous case) due to C/H being depleted further due to CO beyond 2 au and CH$_4$ between 1 and 16 au in gas phase being destroyed by increased ionization and the resulting grain surface reaction products CO$_2$, H$_2$CO and C$_2$H$_6$ being frozen out in the solid phase \citep{eistrup2016setting}. This also has the effect of reducing gas phase O/H abundance as well. At 20 au, the overall C/H and O/H abundances are even more significantly depleted due to freezing out of CH$_4$ and O$_2$ (iceline at 21 au) in addition to the depletion causes already mentioned above. The dichotomy between the values of N/H at 0.5-1 au and 5-20 au for both cases is because the NH$_3$ iceline lies at 2.5 au. But, N$_2$ in the gaseous phase is still more than enough for all the values to remain similar within the same order of magnitude.
\\
\\
The C/O ratio behaviour is very different for the two sub-cases of atomic initial abundances in Table 1. Since in this case the disc chemistry starts from atoms and the formation of gas phase molecules at a particular location depends on its temperature and ionization conditions. At radius $>$ 0.3 au, CO and O$_2$ are the major C and O carrying molecules and N$_2$ is the major N carrying molecule. CH$_4$ is almost completely depleted compared to the inheritance scenario. This is the reason why the C/O ratio generally remains sub-solar for all the locations considered. At 5 au, C/O ratio falls even further because of the CO to CO$_2$ ice conversion mechanism pointed out above, depleting gas phase C/H. The C/O is highest at 20 au because here the gas phase O$_2$ starts freezing out.
\\
\\
All 32 input values (16 for each P-T profile) for elemental abundances with model labels, location, presence of an inversion in the P-T profile, ionization and C/O ratio details are summarized in Table \ref{table:1}. Each set of 3 abundances (C/H, O/H and N/H) are assumed to be the starting initial elemental abundances for the gas accreted onto the envelope of HD 209458b and then fed into VULCAN as initial elemental abundances in order to evolve the chemistry of its atmosphere. 
\subsection{JWST observation simulation}
JWST observations were simulated using the GUI version of the open source PandExo \citep{batalha2017pandexo} package using a similar procedure as provided in \citet{zilinskas2020atmospheric}. NIRISS cannot be used for HD 209458 as the star is brighter than the lower observing limit for saturation for this instrument i.e. J band magnitude (in near infrared transmission window) = 6.6 \citep{wenger2000simbad} versus recommended lower limits of 7 for Substrip 96 and 8 for Substrip 256 respectively. We also do not use NIRSpec as HD 209458 is brighter than what can be observed (without saturation) using its prism mode and its brightness is close to the upper limit of brightness that can be observed using the grism modes i.e. J band magnitude = 6.6 versus a recommended lower limit of 6. NIRCam grism's F322W2 and F444W (using SUBGRISM64 which has the fastest readout) and MIRI LRS (slitless) modes are used to simulate observations over 2.4 $\mu$m to \textbf{13} $\mu$m because the brightness of our source falls safely within the limits of both instruments i.e. K value of 6.3 \citep{wenger2000simbad} vs a lower limit of saturation of 4 for MIRI. Noise levels of 30 ppm and 50 ppm were used for both respectively \citep{zilinskas2020atmospheric}. A saturation limit of 80\% of full well capacity was specified and we chose four transits in total to simulate observations. These observations were later binned to a spectral resolution of R=100 for 4 transits similar to \citet{zilinskas2020atmospheric} (but who used R=15). Both modes of NIRcam grisms we have considered have an overlapping range of observation wavelengths with different sensitivities. Hence, we have retained the observations of F322W2 from 2.4-4 $\mu$m and of F444W from 4-5 $\mu$m respectively to capture the best results from these instruments.

\section{Results}
\begin{figure*}
    \centering
    \includegraphics[width = 2\columnwidth]{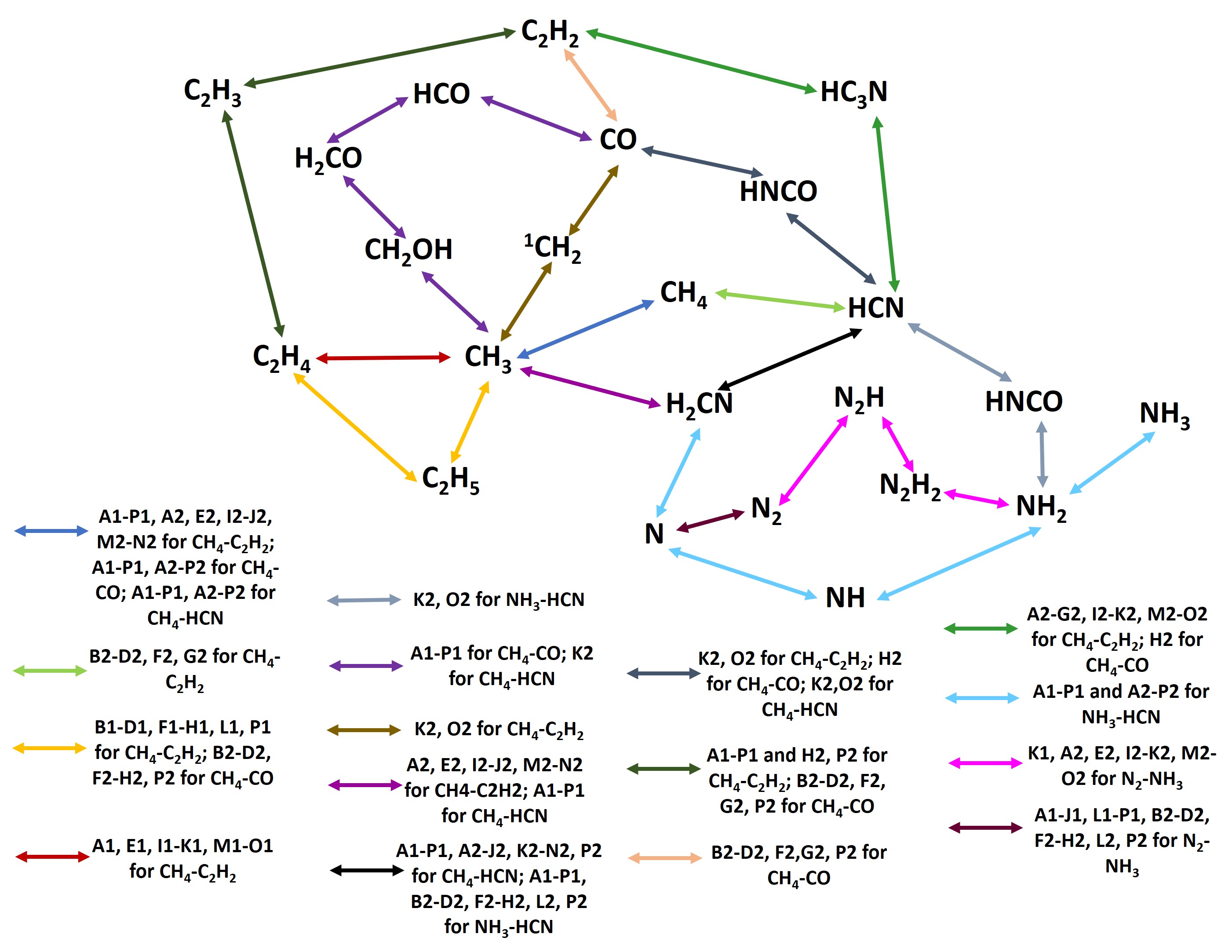}
    \caption{Visualization of major chemical pathways between a few selected molecules. All of these occur at P = 1 mbar. Note that HNCO has been depicted twice for better clarity. For individual model-wise reaction pathway analysis with addition of all secondary reactants, please see the Appendix.}
    \label{fig:sCN}
\end{figure*}
We compare the atmospheric mixing ratios of all volatiles mentioned in Section 2.2 for the differing cases of molecular vs. atomic initial abundances, levels of disc ionization (low or high) and P-T profiles (with and without a thermal inversion). We leave out H$_2$ and He as they dominate in all cases but have little effect on the spectra beyond Rayleigh scattering and Collision Induced Absorption (CIA). Following that, we then compare the differences in synthetic transmission and emission spectra for all these cases.
\\
\\
For aiding the analysis of the chemistry plots obtained from our models, we use the prescription provided in \citet{tsai2018toward} which used Djikstra's algorithm in order to find the shortest path and the Rate Limiting Step (RLS) for a reaction converting one species to the other (see Appendix B of \citet{tsai2018toward}). We construct schematic plots for chemical pathways between CH$_4$-CO, CH$_4$-C$_2$H$_2$, CH$_4$-HCN, NH$_3$-HCN and N$_2$-NH$_3$ at P = 1 mbar. Our choice of the pressure level was motivated by the fact that this pressure is close to the reference photospheric level we assume to probe the atmosphere and as it also represents the location where the non-inverted P-T profile starts being cooler than the inverted P-T profile. Types of reaction pathways and the model-wise distribution of such pathways are covered in the Appendix. This allows us to build comprehensive networks of chemical pathways connecting CH$_4$, CO and HCN and N$_2$, NH$_3$ and HCN which we show in Figure \ref{fig:sCN} and is the main plot for our pathway analysis below. Pathways are coloured differently depending on their model and specific network dependence. 
\subsection{Trends in atmospheric mixing ratios of major volatiles under different elemental abundances}
\subsubsection{For molecular initial abundances}
\begin{figure*}
    \centering
    \includegraphics[width = 2\columnwidth]{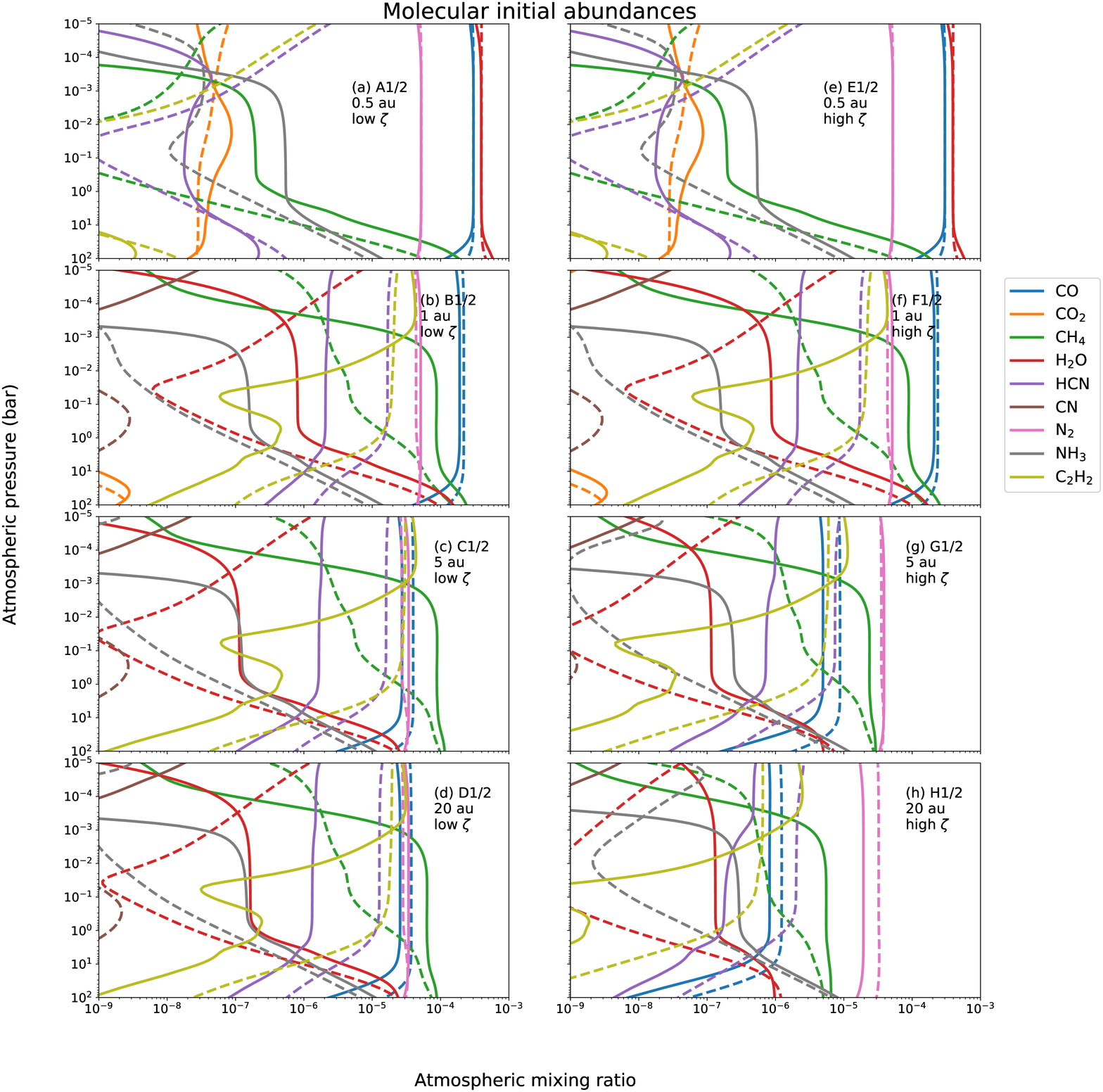}
    \caption{Mixing ratios for specific volatiles (CO, CO$_2$, CH$_4$, H$_2$O, N$_2$, HCN, NH$_3$ and C$_2$H$_2$) in the atmosphere for HD 209458b for different positions (0.5 au, 1 au, 5 au and 20 au) of local gas accretion and varying levels of disc ionization ($\zeta$) in the case for molecular initial abundances (inheritance scenario). Solid lines are for an inverted P-T profile and dotted lines are for a non-inverted P-T profile. Chemical evolution is assumed to happen at the exoplanet's current location.}
    \label{fig:1}
\end{figure*}

For the case of molecular initial abundances with low ionization with an atmospheric temperature inversion shown in panels a, b, c and d of Figure \ref{fig:1} (mixing ratios in solid lines for an inverted P-T profile) corresponding to Models A1 to D1, the dominant volatile (apart from H$_2$ and He) in the atmosphere goes from H$_2$O for A1 to CO for B1 and then CH$_4$ for C1 and D1. Moving upwards through the atmosphere (from 100 bar), the mixing ratio of H$_2$O typically falls before quenching at around a pressure level of 10 bars (for A1) or 1 bar (for B1 to D1) before it starts falling off in the upper atmosphere again ($<$ 10$^{-3}$ bar). The mixing ratio of H$_2$O at each level is set primarily by the balance between the destruction reactions: H+H$_2$O$\rightarrow$OH+H$_2$ and M+H$_2$O$\rightarrow$M+H+OH (where M is any other different species), and the formation reactions which are the reverse of these reactions. Both are fast for the case of a hot planet like HD 209458b and hence the mixing ratios for H$_2$O remain close to equilibrium until it is quenched in the lower atmosphere and this extends until millibar levels at which point destruction of of H$_2$O starts being important as temperature exceeds ~1900K.
\\
\\
CO is the second most dominant volatile for Panel a (A1) and third dominant (after CH$_4$ and N$_2$) for Panels c and d (C1 and D1) and the most dominant volatile for Panel b (Model B1). The mixing ratio for CO increases as we move upwards in the lower atmosphere before it quenches at around 10 bars and remains at that level. The mixing ratio for CO at each level is determined by a balance between its destruction primarily by the reactions: H+CO+M $\rightarrow$ HCO+M and CO+H$_2$ $\rightarrow$ HCO + H, and its formation by the reverse of these reactions. The first reaction is the major pathway in the CO-CH$_4$ interconversion for these atmospheric models indicated in Figure \ref{fig:sCN}. For a hot planet like HD 209458b, both the forward and backward reactions are fast enough so that CO remains close to its equilibrium mixing ratio values and is quenched pretty low in the atmosphere and continues at that ratio until about micro bar levels. The levels of both CO and H$_2$O in turn determine the mixing ratio of CO$_2$ in the atmosphere with it being in a pseudo-equilibrium with both these species \citep{moses2011disequilibrium}. Hence, as H$_2$O becomes depleted with increase in C/O, the mixing ratio of CO$_2$ also falls off accordingly. As would be apparent in Figure \ref{fig:sCN}, CO is also responsible for the levels of CH$_4$ and HCN in the atmosphere through the interconversion pathways. Hence, with increase in C/O, levels of both these species also increase.
\\
\\
The abundances of both H$_2$O and CO depend on the C/O ratio, with H$_2$O generally decreasing and CO generally increasing with the increase in C/O ratio. For C/O $<$ 1, CO is generally the major C carrying species and H$_2$O is the major O carrying species. At a C/O = 0.5, these species have almost equal mixing ratios. For a sub-solar C/O ratio, C would be the limiting element and the abundance of CO would slightly decrease. With a super-solar C/O ratio, the availability of C becomes higher and hence CO would typically increase in mixing ratio unless C is already heavily depleted in absolute abundance (panels c and d or Models C1 and D1). With C/O $>$ 1, O becomes the limiting element and almost all O now ends up in CO with H$_2$O being depleted. 
\\
\\
The mixing ratio for CH$_4$ starts decreasing in the lower atmosphere before quenching in the mid-atmosphere (1 bar - 1 mbar) before falling again in the upper atmosphere ($<$1 mbar). The mixing ratio at which CH$_4$ quenches increases with the C/O ratio due to an increased availability of C. The fall of CH$_4$ levels above 1 mbar is due to increased prominence of the two reaction pathways forming HCN (CH$_4$-HCN pathway for these models) and C$_2$H$_2$ by either CH$_4$-CH$_3$-C$_2$H$_4$-C$_2$H$_3$-C$_2$H$_2$ for low C/O ratios (A1) or CH$_4$-CH$_3$-C$_2$H$_5$-C$_2$H$_4$-C$_2$H$_3$-C$_2$H$_2$ for high C/O ratios (B1-D1) respectively (Figure \ref{fig:sCN}).
\\
\\
N$_2$ and NH$_3$ are very insensitive to changes in all panels since they do not have any C or O and the N/H values do not differ a lot for our input values and also because both of them are significant carriers of N. However, while N$_2$ has a constant mixing ratio for almost all pressure levels, NH$_3$ shows a behaviour of fall, quench and fall for its profile similar to that of CH$_4$ since both NH$_3$ and CH$_4$ undergo kinetic conversion reactions to N$_2$ and CO respectively (see Figure \ref{fig:sCN}). N$_2$ and CO in turn are both major species in the atmosphere and in our cases are quenched for most of the atmosphere. In addition, NH$_3$ is an essential molecule for producing HCN in all models using the NH$_3$-HCN pathway. Thus, NH$_3$ is still indirectly but weakly dependent on the C/O ratio. This behaviour for all species discussed until now is expected and is seen in other disequilibrium models \citep{moses2011disequilibrium, venot2012chemical, hobbs2019chemical} for this planet.
\\
\\
The mixing ratio of HCN gradually rises with increasing C/O ratio since the amount of carbon available to react with nitrogen increases and is highest at B1-D1. The behaviour of HCN is, however, more complex than the other species we have considered and is determined by the quench and equilibrium levels of NH$_3$, CH$_4$ and H$_2$ as seen in the profile discussions for CH$_4$ and NH$_3$ above. Apart from A1 where HCN never manages to find a quench level because its production is too inefficient, B1-D1, with high C/O ratios, have enough C (through CH$_3$) for HCN production to be fast resulting in high enough abundances to be quenched upwards of ~1 bar.
\\
\\
The mixing ratio of C$_2$H$_2$ is overall pretty low in the lower atmosphere in all panels but it becomes prominent for the mid-atmosphere and upper atmosphere as the C/O ratio is increased. For low C/O ratios (A1), C$_2$H$_2$ is produced by CH$_4$-CH$_3$-C$_2$H$_4$-C$_2$H$_3$-C$_2$H$_2$ where the CH$_3$-C$_2$H$_4$ step is a slow rate limiting step mediated by the excited singlet $^1$CH$_2$. C$_2$H$_2$ manages to reach a stable mixing ratio value (but not an actual quench level) at around 1 mbar only for super-solar C/O ratios (B1-D1), where it proceeds by the pathway CH$_4$-CH$_3$-C$_2$H$_5$-C$_2$H$_4$-C$_2$H$_3-$C$_2$H$_2$. The rate-limiting step is the formation of C$_2$H$_5$ by reaction of two CH$_3$ radicals. The abundance of CH$_3$ is increased high up in the atmosphere by the dissociation of CH$_4$. The rate of C$_2$H$_5$ formation is about 3 times faster in this model compared to A1.
\\
\\
CO$_2$ also becomes more prominent only at lower C/O ratios where it can compete with CO and overall its abundance is very sensitive to the C/O ratio due to O becoming a limiting element as C/O increases. CN, while included here, doesn't show up in the lower atmosphere in any of the plots but does appear in the atmosphere above 0.1 mbar for super solar C/O ratios as dissociation of HCN becomes viable. This sensitivity of CH$_4$, HCN, C$_2$H$_2$ and CO$_2$ to C/O ratio are what is already expected from results of disequilibrium chemical kinetics \citep{moses2014chemical}.
\\
\\
For the case of molecular initial abundances with high ionization with a temperature inversion  shown in panels e, f, g and h in Figure \ref{fig:1} (solid lines) corresponding to Models E1 to H1, the dominant volatile (apart from H$_2$ and He) depends on the assembly location in a similar way to the low ionization case. For E1, the atmosphere is dominated by H$_2$O (due to the sub-solar C/O ratio at this location), for F1 it is CO and for G1 and H1 the main molecule is N$_2$. The difference in the main volatile between C1-D1 compared to G1-H1 is due to an overall depletion in absolute abundance of C/H by about a magnitude for the high ionization cases. In comparison, the N/H abundances for the high ionization cases remain almost same as with the low ionization cases - this is illustrated by the reduced C/N ratios in Table \ref{table:1}). 
\\
\\
The mixing ratio for CH$_4$ steadily goes up with the C/O ratio (as already explained above) and it becomes the second dominant volatile in F1, G1 and H1. N$_2$ and NH$_3$ remain insensitive to changes in all panels as absolute N/H abundances do not change much. In general, the mixing ratio profile behaviour for models E1 and F1 are same as their low ionization counterparts (models A1 and B1), but models G1 and H1 differ from what we might expect based on their low ionization counterparts (Model D1 based on C/O ratio). Model G1 has a slightly lower C/O ratio than D1 but with much lower initial absolute C/H and O/H abundances which causes the overall shift of all profiles with C and O to the left. Even though D1 and H1 have the same C/O ratios, the effect of increased cosmic ray ionization on chemistry is very clear, as the mixing ratios of all C and O bearing species are much lower in H1 compared to D1. This is because the initial abundance of both C/H and O/H is an order of magnitude lower for H1 (values are even lower than the level for G1) in comparison to D1. HCN in H1 has a more complicated behaviour, with two quench levels instead of the one seen in C1 and D1. One quench level is at the location of the CO-NH$_3$ quench level (in the lower atmosphere at around 1 bar) and the other is at the location where both CH$_4$ and NH$_3$ abundances start to fall (in the upper atmosphere at  around 1 mbar). The absolute C/H depletion might be the cause of this profile behaviour as well considering that the reaction pathway is the same as in previous cases.
\\
\\
For the case of molecular initial abundances with low ionization and no temperature inversion  shown in panels a, b, c and d in Figure \ref{fig:1} (dotted lines for a non inverted P-T profile) corresponding to Models A2 to D2, the dominant volatile (apart from H$_2$ and He) goes from H$_2$O (sub solar C/O ratios) in A2 to (mostly) CO in the rest of the three panels (super solar C/O ratios). Unlike the case with a temperature inversion, CH$_4$ is never a prominent volatile in the mid-atmosphere and upper atmosphere as in the temperature inversion case but does increase in mixing ratio level as C/O increases. This is due to several other reaction pathways in the CO-CH$_4$ interconversion (see Figure \ref{fig:sCN}) becoming possible in the lower and mid-atmosphere due to the non-inverted P-T profile being considerable hotter than the inverted P-T profile for much of the atmosphere below the 1 mbar level. However, above ~ 1 mbar level, the mixing ratio of CH$_4$ is larger compared to the case of an inverted profile because the non-inverted profile is now cooler than the inverted profile which reduces the rate of its dissociation.
\\
\\
N$_2$ remains very insensitive to the C/O ratio as in all other cases. H$_2$O and hence CO$_2$ on the other hand remain very sensitive to C/O. However, the behaviour of H$_2$O is different in comparison to inverted profile cases as its mixing ratio decreases more rapidly but then increases just above the 0.1 bar level. This level coincides with the pressure level above which the inverted P-T profile starts exhibiting its inversion by heating up while the non-inverted profile continues to cool down. So, while dissociation of water starts getting enhanced for the inverted profile (even more after it cools down from ~1900K), the formation reaction rates are enhanced for the non-inverted profile going below that temperature, resulting in the profile difference seen here.
\\
\\
NH$_3$ is less prominent than in cases without a temperature inversion as most of it either quickly reacts with CH$_3$ (from CH$_4$) for B2-D2 (high C/O ratio) to form HCN through the same NH$_3$-HCN pathway mentioned before. The kinetic interconversion of NH$_3$ and N$_2$ is also not fast enough for NH$_3$ to reach a quenching level since the rate limiting steps (N$_2$-N and N$_2$H-N$_2$H$_2$) of both chemical pathways responsible (see Figure \ref{fig:sCN}) are very slow.
\\
\\
HCN quenches at higher mixing ratios at the same pressure levels compared to the cases with a temperature inversion for B2-D2. In A2, HCN becomes very prominent in the upper atmosphere. This is because HCN, in addition to being the product of faster NH$_3$-HCN and CH$_4$-HCN pathways mentioned before, is also a major step in all three CH$_4$-C$_2$H$_2$ pathways possible in these cases. The conversion from HCN to HC$_3$N (followed by a conversion to C$_2$H$_2$) mediated by $^1$C$_2$H is a rate limiting step in all these pathways. Since C$_2$H$_2$ is correlated with HCN, the mixing ratio profiles for both molecules behave similarly in all models. In addition, C$_2$H$_2$ is also formed by one pathway in the CH$_4$-CO interconversion process. So, the level of C$_2$H$_2$ is sometimes slightly higher than the level of HCN in mid and upper atmospheres. However, HCN is more spectrally significant due to its higher opacity which often  overshadows that of C$_2$H$_2$ in the corresponding transmission spectrum plots.
\\
\begin{figure*}
    \centering
    \includegraphics[width = 2\columnwidth]{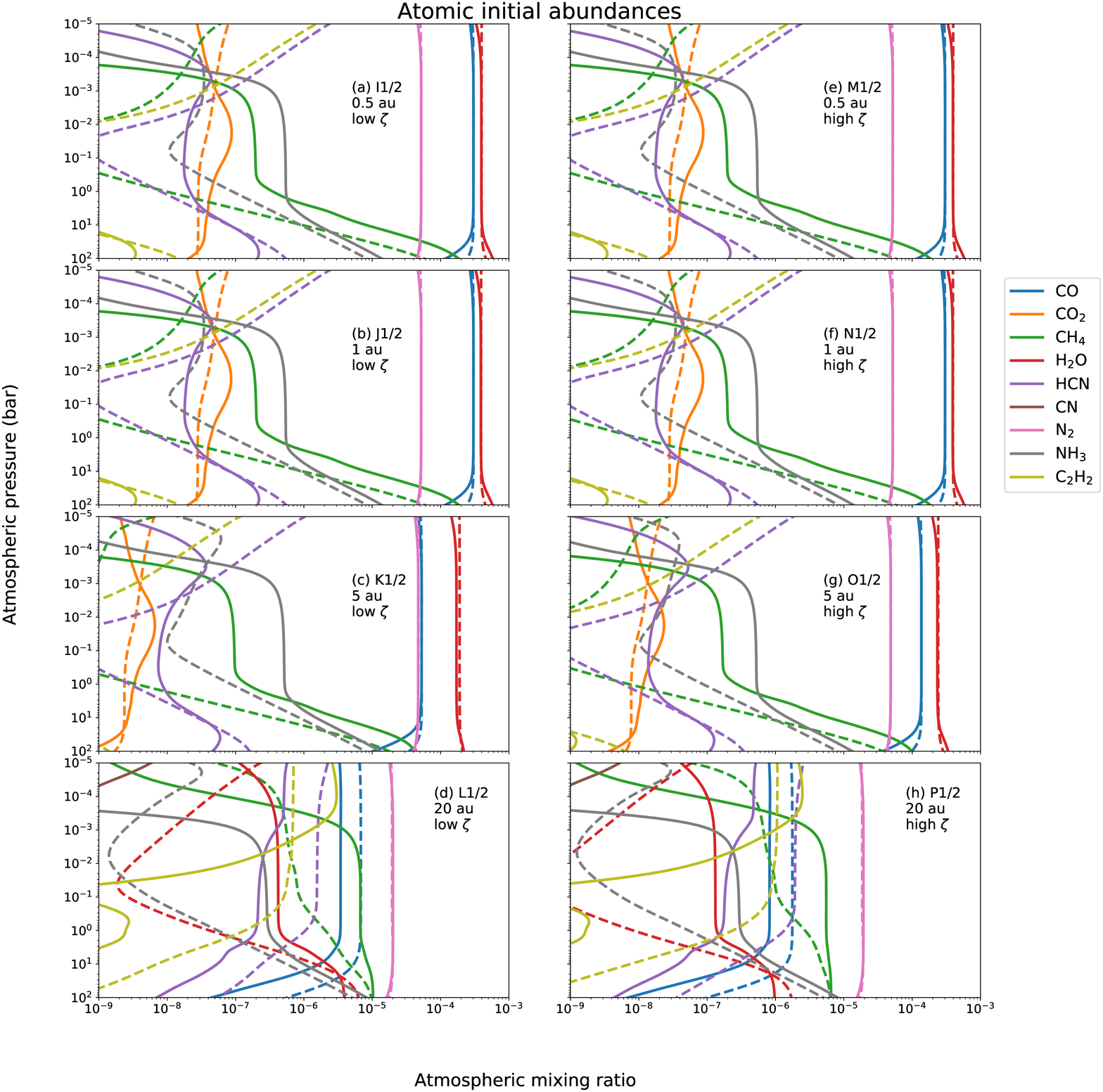}
    \caption{Mixing ratios for specific volatiles (CO, CO$_2$, CH$_4$, H$_2$O, N$_2$, HCN, NH$_3$ and C$_2$H$_2$) in the atmosphere for HD 209458b for different positions (0.5 au, 1 au, 5 au and 20 au) of local gas accretion and varying levels of disc ionization ($\zeta$) in the case for atomic initial abundances (reset scenario). Solid lines are for an inverted P-T profile and dotted lines are for a non-inverted P-T profile. Chemical evolution is assumed to happen at the exoplanet's current location.}
    \label{fig:2}
\end{figure*}
\\
For the case of molecular initial abundances with high ionization and no temperature inversion  shown in panels e, f, g and h in Figure \ref{fig:1} (dotted lines) corresponding to Models E2 to H2, the behaviour of major volatiles is the same as in the case of low ionization. Models E2 and F2 have similar behaviour as A2 and B2 which have similar C/O ratios. However, H2 doesn't have a similar profile behaviour as D2 even though both have similar C/O ratios because its C/H and O/H are more than an order of magnitude less in comparison to D2. This causes almost all of the mixing ratio profiles to shift to the left.
\\
\\
The difference in the solid mixing ratio profiles between G1 and H1 with respect to D1 and between the dotted mixing ratio profiles H2 with respect to D2 for Figure \ref{fig:1} shows that it is not just the C/O ratio of the gas but also the overall absolute abundances that affect the chemistry in atmospheres. This absolute abundance value is set by the level of chemical processing in the disc. This overall abundance, however, is reflected in changes in both C/O and C/N ratios together as seen in Table \ref{table:1}. Hence, both the C/O and C/N ratios taken together can provide a more accurate description. This was the basis behind \citet{notsu2020composition} and \citet{turrini2021tracing} where they used the O/H ratio in addition to the C/O ratio, and the C/O and C/N ratios together respectively to comment on formation locations.

\subsubsection{For atomic initial abundances}
\begin{deluxetable*}{cc}
\tablewidth{0pt}
\tablecaption{Features exhibited by few spectrally significant molecules \label{table:2}}
\tablehead{
\colhead{Molecule} & \colhead{Spectral features ($\mu$m)}
}
\startdata
H$_2$O & 0.7-0.8, 0.8-0.9, 1, 3 peaks between 1-2, 2-4, 5, broad features at 7-9 and 10 onward  \\
CH$_4$ & 0.9, 1, 3 peaks between 1-2, 2.5, 3.5 and broad feature from 6-9  \\
NH$_3$ & 2-3 feature, 10 and broad feature from 10-25 \\
CO and CO$_2$ & 4.1 and 4, 14 \\
HCN & 1.4, 1.5, 1.8, 2.6, 2.9, 3.1, 3.6, 4, 5, 7-9, 10, 13 and 20 \\
CN & 0.3-0.5 and 0.7-1.1 (Sharp jagged peaks) \\
C$_2$H$_2$ & features from 1.5-2, 3-4, 7-9, peak at 14\\
H$_2$ and He Rayleigh scattering & Negative slope from 0.3-1.2 \\
\enddata
\end{deluxetable*}

\begin{figure*}
    \centering
    \includegraphics[width = 2\columnwidth]{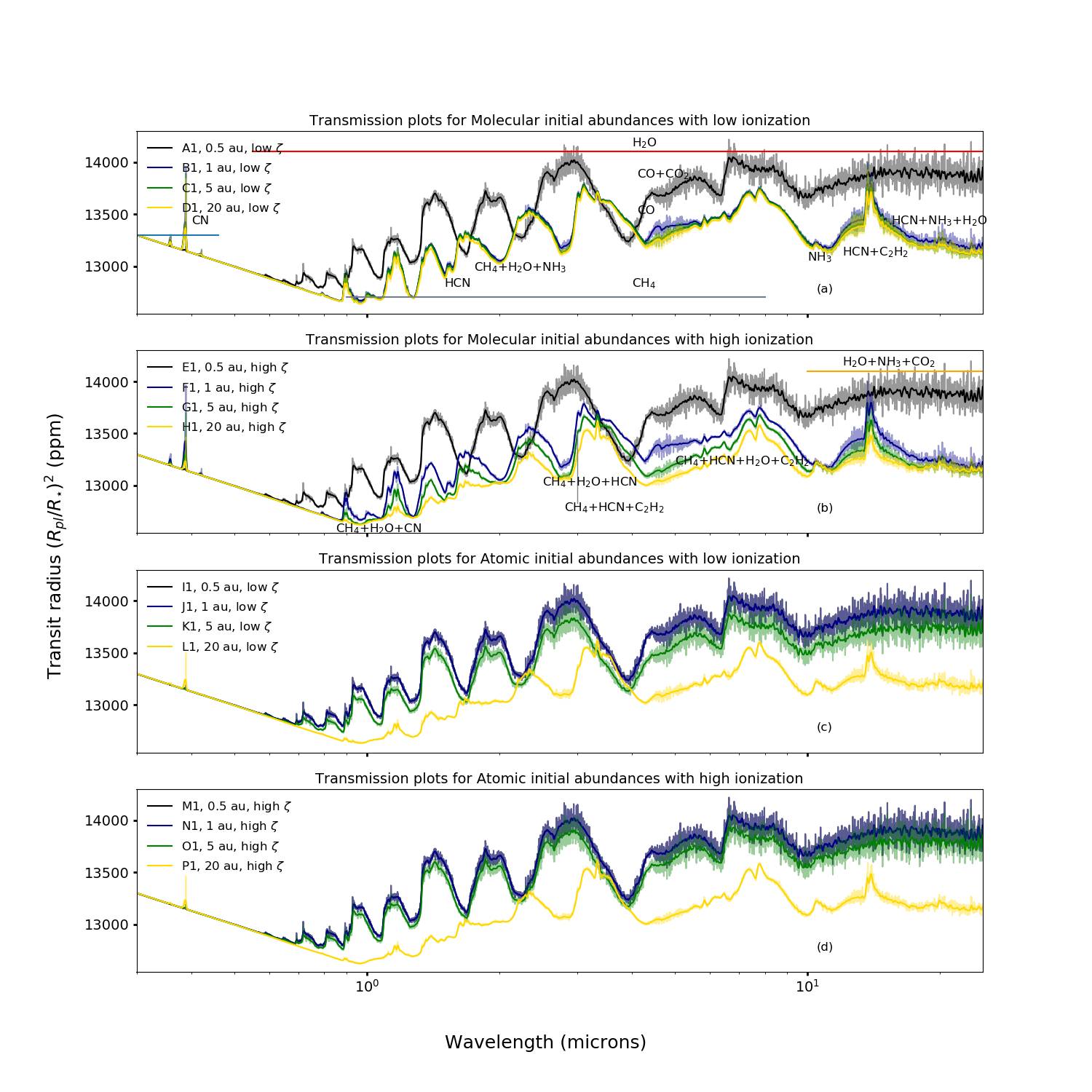}
    \caption{Synthetic transmission spectra for all cases of abundances considered for chemistry in HD 209458b's atmosphere with the presence of a thermal inversion. The peaks for H$_2$O and CH$_4$ are multiple and span the entirety of red and gray horizontal lines (a signature of any other species falling in between is specifically labelled). The faded spectra at the back have a R=1000 while the highlighted spectra for easy visibility are lower resolution spectra obtained by convolving the original spectra with a Gaussian kernel of standard deviation=3 following a prescription suggested in \citet{zilinskas2020atmospheric}.}
    \label{fig:3}
\end{figure*}

For the case of atomic initial abundances with low ionization with a temperature inversion  shown in panels a, b, c and d (solid lines) in Figure \ref{fig:2} corresponding to Models I1 to L1, H$_2$O and CO are the major volatiles apart from H$_2$ and He) in the atmosphere for I1-K1 (sub solar C/O ratios) whereas N$_2$ is the dominant volatile for L1. Even though the C/O for L1 is closer to I1 and J1, the overall initial C and O abundances with respect to H are about an order of magnitude lower, while N/H mostly remains the same and this is reflected in the C/N ratios. The depletion causes mixing ratio profiles in L1 to behave like models with high C/O ratios (for thermally inverted profile). This is also reflected in the reaction pathways with the one for CH$_4$-C$_2$H$_2$ being the same one as in high C/O models with thermally inverted profile. Other than the anomalous behaviour of L1, all other models behave as expected based on their C/O ratios as mentioned in cases before.
\\
\\
Models with atomic initial abundances, high ionization and a temperature inversion are shown in panels e, f, g and h in Figure \ref{fig:2} (solid lines) corresponding to Models M1 to P1. H$_2$O and CO apart from H$_2$ and He) are again the dominant volatiles in most cases (sub-solar CO ratios) except for P1 which has a super-solar C/O and where N$_2$ and CH$_4$ are dominant. N$_2$ and NH$_3$ continue to be insensitive for the same reasons as pointed out before. All models behave as expected based on their C/O ratios.
\\
\\
For the case of atomic initial abundances with low ionization with no temperature inversion  shown in panels a, b, c and d in Figure \ref{fig:2} (dotted lines) corresponding to Models I2 to L2, H$_2$O apart from H$_2$O and He) is the predominant volatile (sub-solar C/O ratios) followed by CO except for L2 where N$_2$ is the dominant volatile and followed by CO. All models, except L2, behave as expected based on their C/O ratios. However, the chemical pathways for these models do sometimes differ from their molecular abundance counterparts. The case for L2 as an anomalous case was also made in the case of a thermally inverted profile as well (model L1). So, the abundances used for such models (L1/2) can cause the transmission or emission spectrum to behave as one for a high C/O ratio even though the actual C/O ratio would be sub-solar.
\\
\\
For the case of atomic initial abundances with high ionization with no temperature inversion  shown in panels e, f, g and h in Figure \ref{fig:2} (dotted lines) corresponding to Models M2 to P2, H$_2$O apart from H$_2$ and He) is almost always dominant with CO following it (sub-solar C/O ratios), except for P2 which has a super-solar C/O ratio and behaves similar to L2 and other models with high C/O ratios. All models behave as expected based on their C/O ratios.

\subsection{Synthetic transmission and emission spectra}
\subsubsection{Transmission spectra}
\begin{figure*}
    \centering
    \includegraphics[width = 2\columnwidth]{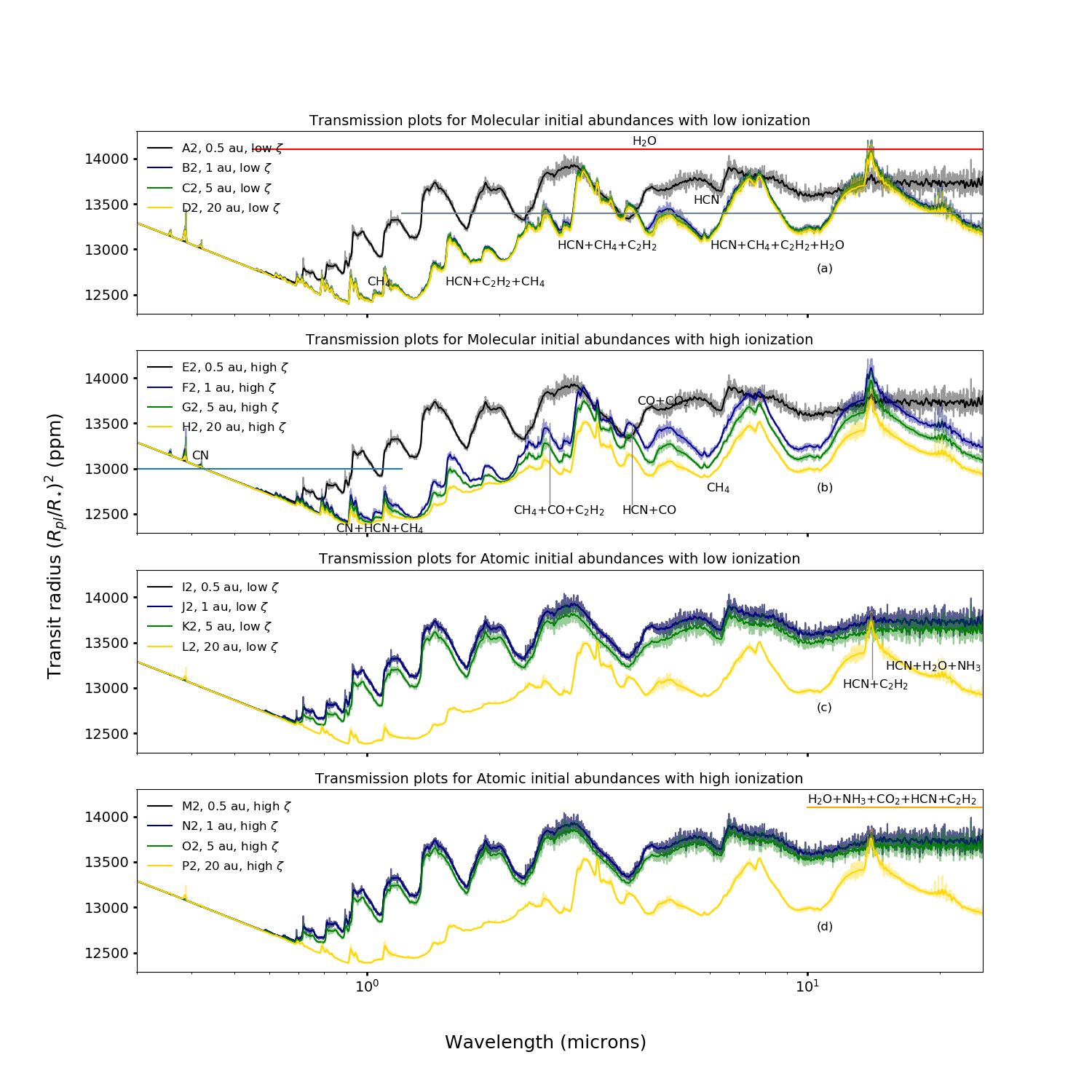}
    \caption{Same as Figure \ref{fig:3} but for a P-T profile without thermal inversion. In addition, CN peaks span the blue horizontal line.}
    \label{fig:7}
\end{figure*}

The transmission spectra are shown in Figure \ref{fig:3} for cases with a temperature inversion (Models A1 to P1) and in Figure \ref{fig:7} for cases with no temperature inversion (Models A2 to P2). The absorption spectral features of H$_2$O, CO, CO$_2$, HCN, CN, CH$_4$, C$_2$H$_2$ and NH$_3$ are summarized in Table \ref{table:2}. Accordingly, all panels in Figure \ref{fig:3} and Figure \ref{fig:7} can be divided into three types of spectra: majority H$_2$O, majority CH$_4$ and majority CN-HCN.
\\
\\
Figure \ref{fig:3}a shows the spectra generated for models with a temperature inversion, low ionization and molecular initial abundances (models A1 to D1). The molecules that generally dominate the spectra correspond to those with high abundances in such models in Figure \ref{fig:1}. B1, C1 and D1 have very similar spectra, with prominent CH$_4$ (mixed with small HCN, H$_2$O and NH$_3$) signatures in the 1-10 $\mu$m range, followed by a small NH$_3$ peak at 10 $\mu$m, followed by a significant HCN+C$_2$H$_2$ peak at 13-14 $\mu$m and ending with an almost flat spectrum from 15-25 $\mu$m due to the presence of HCN, NH$_3$ and H$_2$O. There is also a small peak at 4.1 $\mu$m corresponding to CO. For A1, the predicted spectrum is almost entirely due to H$_2$O which dominates over all other features in the 0.6-25 $\mu$m range. The concave shaped 10-25 $\mu$m spectrum while being shaped by H$_2$O also has contributions from NH$_3$ and CO$_2$ as well. A1 also has a small broad peak at 4-4.1 $\mu$m due to the presence of both CO$_2$ and CO respectively. All the models however show prominent CN features at 0.3 $\mu$m. Considering that CN is not very abundant in the atmospheres in such models, this is made possible due to the large opacity of CN at such wavelengths.
\\
\\
In Figure \ref{fig:3}b we can see that Models E1 to H1 (similar to models A1 to D1 but with high ionization instead) result in  similar spectral features as the previous models, and these can also be explained using the same logic as above. Hence, for the complete case of initial molecular abundances with a temperature inversion, a super-solar C/O ratio with HCN, CH$_4$, H$_2$O, CO, CN and NH$_3$ signatures would indicate gas accretion anywhere between 1-20 au, i.e. between the H$_2$O and CO icelines, but a H$_2$O predominant spectrum, with CO$_2$ and CO detection at 4-4.1 $\mu$m, would point to accretion interior to the H$_2$O iceline and a C/O ratio less than solar.
\\
\begin{figure*}
    \centering
    \includegraphics[width = 2\columnwidth]{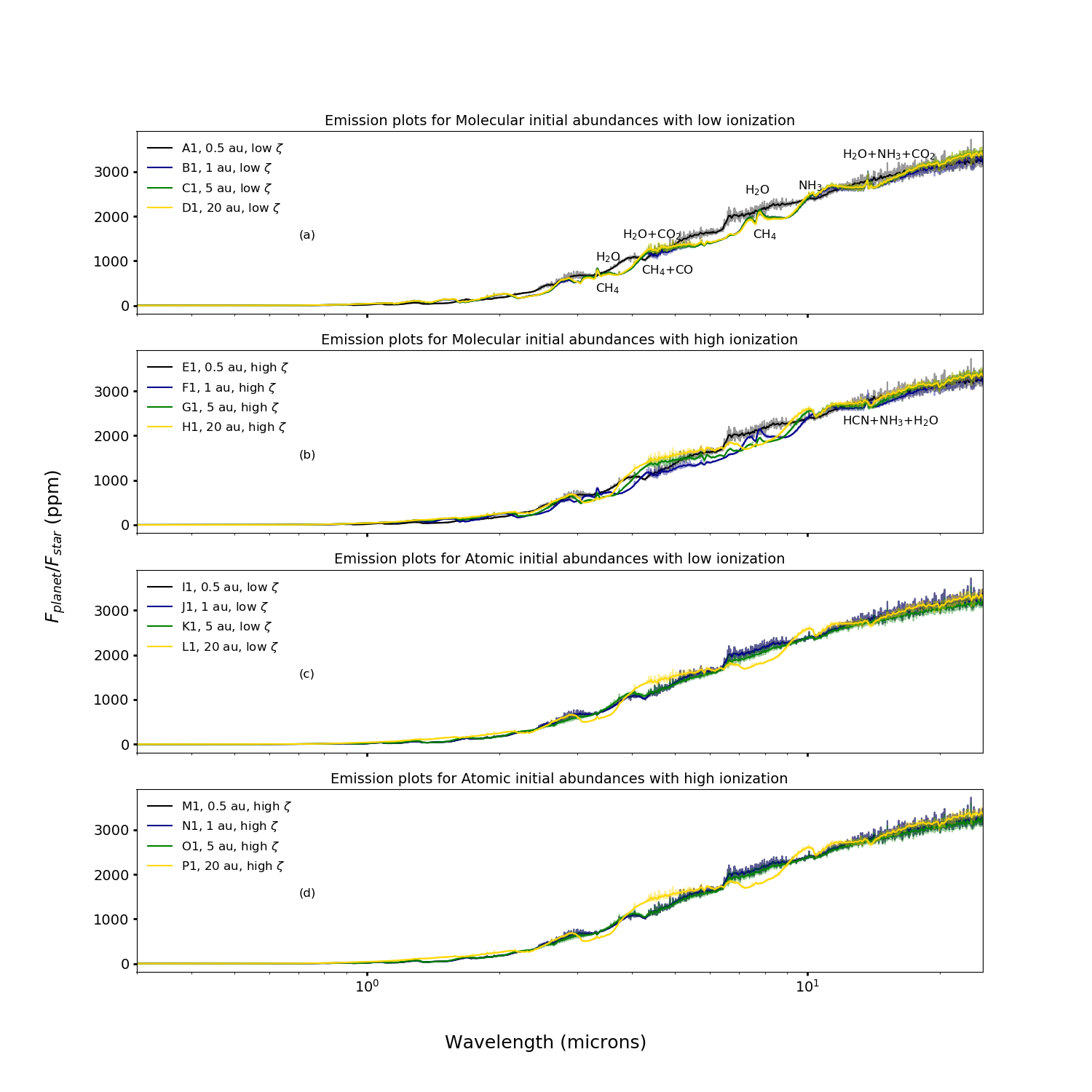}
    \caption{Synthetic emission spectra for all cases of abundances considered for chemistry in HD 209458b's atmosphere with a P-T profile with thermal inversion. The broad peak for H$_2$O at 7-9 $\mu$m and NH$_3$ at 10 $\mu$m are the only ones that can be reasonable discerned. The faded spectra at the back have a R=1000 while the highlighted spectra for easy visibility are lower resolution spectra obtained by convolving the original spectra with a Gaussian kernel of standard deviation=3.}
    \label{fig:4}
\end{figure*}
\\
For Models I1 to L1 and M1 to P1 (atomic initial abundances with a temperature inversion, and low and high ionization respectively), the situation is completely flipped, with H$_2$O predominated spectra dominating the cases (Figure \ref{fig:3}c and d). The only models in these sets that are not dominated by H$_2$O features are for accretion at 20 au (Models L1 and P1), i.e. between the CH$_4$ and CO icelines. Both Models L1 and P1 show features comparable to B1-D1 as mentioned above, but with reduced width and height of the secondary features due to depleted overall abundances leaving the CH$_4$ features more prominent between 1-10 $\mu$m. For L1, this is anomalous as it has a sub-solar C/O ratio. So in the case of atomic initial abundances with a thermal inversion, H$_2$O dominated spectra with CO$_2$ and CO detection at 4-4.1 $\mu$m would point to gas accretion interior to the CH$_4$ iceline for sub-solar C/O ratios in both kinds of ionization conditions. However, a spectrum dominated by CH$_4$ that also includes an HCN+C$_2$H$_2$ signature at 13-14 $\mu$m can indicate accretion between the CH$_4$ and CO icelines for a sub-solar C/O ratio and low ionization and super-solar C/O with high ionization.
\\
\\
For atmospheres with molecular initial abundances and without a thermal inversion (Models A2 to H2), all of the predicted spectra can be grouped into either H$_2$O dominant, or HCN(+CN) dominant (Figure \ref{fig:7}a and b). While H$_2$O completely dominates A2 and E2, there are also small features of CO$_2$ at 4 $\mu$m and contributions from NH$_3$, CO$_2$, HCN and C$_2$H$_2$ in the flat spectrum after 10 $\mu$m. This is in line with the chemistry of these models as shown in Figure \ref{fig:1}. For B2-D2 and F2-H2, HCN has prominent features starting from around 1.1 $\mu$m. There are CO, CH$_4$ and C$_2$H$_2$ features mixed in there as well. However, the CN features are now very prominent and show up in the 0.3-1.1 $\mu$m range. In addition, the HCN-C$_2$H$_2$ peak at 13-14 $\mu$m is even more prominent followed by a dip till 25 $\mu$m with contributions from H$_2$O and NH$_3$ as well. Hence, in the case of molecular abundances without a thermal inversion, a predominant H$_2$O spectrum would indicate accretion within the H$_2$O iceline but a HCN(+CN) dominated spectrum would indicate accretion within the CO$_2$ and CO icelines for both types of possible ionizations.
\\
\begin{figure*}
    \centering
    \includegraphics[width = 2\columnwidth]{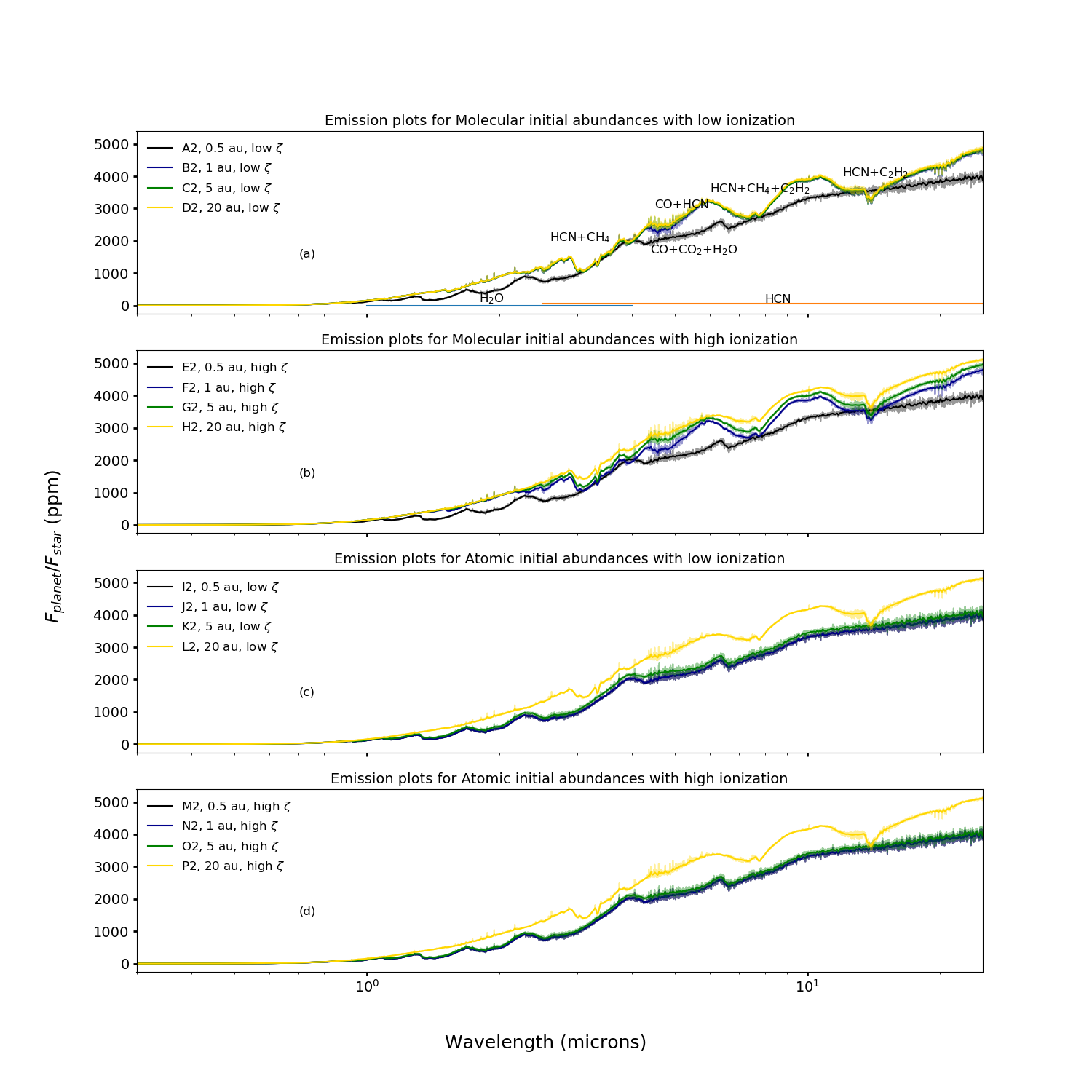}
    \caption{Same as Figure \ref{fig:4} but for a P-T profile without thermal inversion.}
    \label{fig:8}
\end{figure*}
\\
For models I2 to P2 (the case of atomic initial abundances for a non thermal inversion profile (Panels c and d of Figure \ref{fig:7}), H$_2$O dominates almost all the spectra, as is expected based on the chemistry for these particular models (Figure \ref{fig:2}). Hence, H$_2$O would be the dominant spectral signature for gas accretion interior to the CH$_4$ iceline for all such cases. For gas accretion within the CH$_4$ and CO icelines, the spectrum has a similar behaviour to B2-D2 or E2-H2. The anomalous case for L2 being sub-solar but showing super-solar features is similar to L1. Hence, in the case of atomic abundances without a thermal inversion, a predominant H$_2$O spectrum would indicate accretion within the CH$_4$  iceline but a HCN(+CN) dominated spectrum would indicate accretion within the CH$_4$ and CO icelines.

\subsubsection{Emission spectra}
The emission spectra for all models with a temperature inverted profile (Models A1 to P1) are shown in Figure \ref{fig:4} and for a P-T profile without inversion (Models A2 to P2) in Figure \ref{fig:8}. The molecular signatures in both these sets of spectra are noticeably harder to discern from those in the transmission spectra plots which means that observationally it might be preferable to focus on obtaining primary or transit spectra for this planet.
\\
\\
In Figure \ref{fig:4}, the most prominent feature for A1, E1, J1-K1 and M1-O1 are the broad water features at 3-4 $\mu$m, 4-7 $\mu$m and 7-9 $\mu$m. In addition, the small peak for NH$_3$ at 10 $\mu$m can also be slightly differentiated. For B1-D1, F1-H1, L1 and P1, the more difficult to differentiate features include the slight kinks between 3-4 $\mu$m, 4-5 $\mu$m and 8 $\mu$m for CH$_4$. These kinds of mostly featureless emission spectra were also seen in \citet{drummond2016effects} using a P-T profile with thermal inversion for non-equilibrium chemistry (ATMO code) in HD 209458b.
\\
\\
The emission spectra for all abundance values with a non temperature inverted profile are shown in Figure \ref{fig:8}. It can be seen that all these plots have more well defined spectral features compared to the case with thermal inversion. The absorption features for HCN are now prominent for all cases with a HCN dominated spectra (B2-D2, F2-H2, L2 and P2), with multiple troughs ranging between 2.5-25 $\mu$m. For A2, E2, I2-K2 and M2-O2, the H$_2$O absorption features can also be seen. For all the plots, the conclusion remains the same as for the transmission spectra.

\section{Discussion}
\subsection{Comparison with equilibrium plots using TEA}
Our parameters for non-thermally inverted P-T plots for HD 209458b roughly coincide with the parameters for a hot Jupiter accreting gas from the disc with differing initial abundances depending on the position in the disc (as we have also considered) and migrating to 0.05 au in \citet{notsu2020composition} where it evolves its atmospheric chemistry. However, \citet{notsu2020composition} used an analytical P-T profile obtained from \citet{guillot2010radiative}. Thus, it is ripe to compare the effects of disequilibrium over equilibrium as well as any difference resulting from the variations in the P-T profiles on the chemistry of such atmospheres. We henceforth compare the dotted mixing ratio profiles in Figures \ref{fig:1} and \ref{fig:2} to the left plots of Figures 3, 4, 5 and 6 for their paper. 
\\
\\
The most prominent difference among all plots is in the behaviour of HCN. VULCAN disequilibrium chemical kinetics predicts more HCN in the atmosphere in comparison to TEA equilibrium plots in all model cases. The level of HCN predicted by TEA would lead to under-prediction of HCN in both the emission and transmission spectra. Hence, the presence and level of HCN in observed spectra for HD 209458b would be a good predictor of whether the atmospheric chemistry in the atmosphere is in disequilibrium. This is an expected outcome since HCN as a by product of reactions and quenching levels between CH$_4$-NH$_3$ and CO-NH$_3$ and the level of H$_2$ and is expected to show up a lot in hot Jupiters as a notable departure from chemical equilibrium calculations \citep{moses2014chemical, moses2011disequilibrium}. This study validates it for HD 209458b and shows the importance of using disequilibrium chemistry calculations over equilibrium chemistry in a way where it can significantly affect spectral observations. In this way, it remedies one of the limitations that \citet{notsu2020composition}'s paper had already outlined in their work. 
\\
\\
As also expected, C$_2$H$_2$ also shows more prominence in disequilibrium chemistry. But, the level is not enough for it to have a unique spectral fingerprint as it is often overshadowed by the higher opacity of HCN in the same regions. Another difference is the over-prediction of H$_2$O in the lower exoplanet atmosphere but under-prediction in the upper atmosphere by TEA. This can change the shape of the transmission and emission spectra by changing the overall absorption due to H$_2$O in the regions of the atmosphere being probed. The level of CH$_4$ and NH$_3$ is also greater in disequilibrium plots for most cases. This also means that their signatures can now be more prominent in the spectral plots.

\subsection{Examining the presence of thermal inversion in the atmosphere of HD 209458b}
Both our chemistry and spectral results suggest that CH$_4$ is dominant for atmospheric chemistry using a thermally inverted profile at super solar C/O ratios. Consequently, all cases of transmission spectra using such a profile show major fingerprint for CH$_4$. However, a thermally non-inverted profile still shows fingerprints of CH$_4$ but overall is dominated by HCN in both the transmission and emission spectra. Thus, major presence of CH$_4$ in an obtained transmission or emission spectrum for HD 209458b could be a good indicator of whether the temperature profile in the atmosphere is an inverted one similar to \citet{moses2011disequilibrium}, or a non-inverted one similar to the P-T profile used here. Either case would be good indicators of super solar C/O ratios.
\\
\\
However, it must be noted that CH$_4$ detection would only show a difference between the two specific profiles we have considered in this case study. Hence, this claim must not be used as a general fingerprint indicative of a temperature inversion in a hot Jupiter.

\subsection{Limitations of this work}
\begin{figure}
    \centering
    \includegraphics[width = \columnwidth]{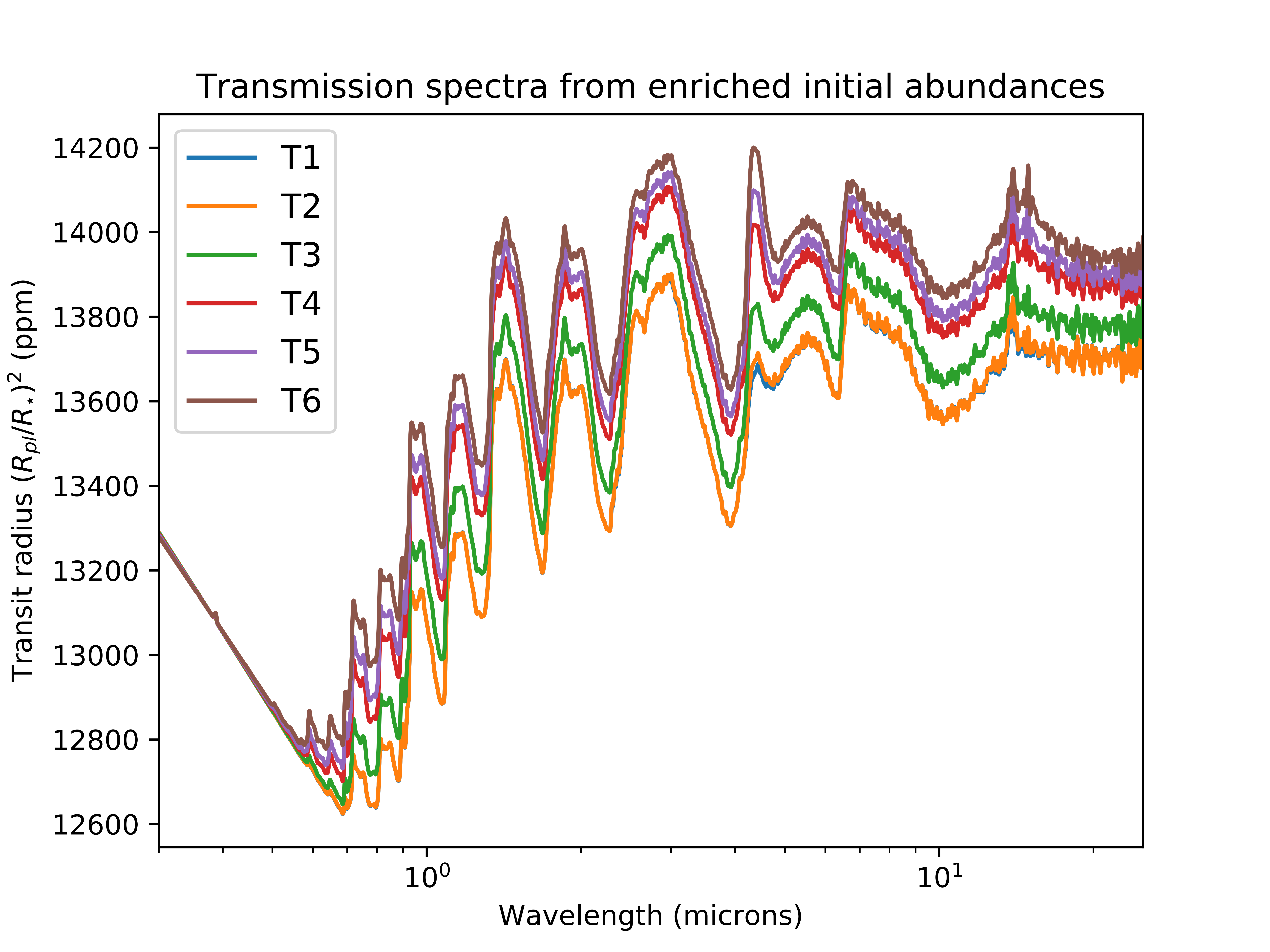}
    \caption{Effect on transmission spectra due to enrichment by solids. Initial elemental abundances are taken from \citet{turrini2021tracing} and the model labels are the same as in Table \ref{table:1}.}
    \label{fig:turrini}
\end{figure}

The first limitation of this work is the limited chemical network used which is a feature of VULCAN. Apart from that, we also don't include sulphur chemistry in this work. However, effects of sulphur chemistry using VULCAN have been discussed in \citet{tsai2021comparative}. \citet{hobbs2021sulfur} recently showed that for the thermally inverted case for HD 209458b, inclusion of sulphur chemistry can reduce the abundance of species like CH$_4$, NH$_3$ and HCN by about two orders of magnitude in between 10$^{-3}$-10$^{-5}$ bar. This might dampen the corresponding spectral features in our spectra. The effect of clouds or aerosols was also not considered in this work. Both of these can also heavily affect the results of this work by damping most of the spectral features as well. Hence, the results of this study must strictly be seen in the case of cloud free hot Jupiter atmospheres only.
\\
\\
Two other limitations from the chemical kinetics model itself is that VULCAN is a 1D model and is not self consistent. While the P-T-Kzz profiles obtained from \citet{moses2011disequilibrium} and HELIOS are definite upgrades compared to using just a simple analytical P-T model based on \citet{guillot2010radiative} (used for all cases in \citet{notsu2020composition}), they still don't take into account the coupling between P-T profiles and the atmospheric chemistry evolution. In addition, 1D models assume a spherically symmetric case for the atmosphere and hence have a reduced computational cost. Such simplifications however neglect the more complicated 3D behaviour of heterogeneous exoplanet atmospheres which can affect the abundance of elements in the atmosphere and affect the overall analysis of this paper.
\\
\\
Limitations of the disc chemistry model used in this analysis arise from the fact that it follows chemical evolution for a static protoplanetary disc within 1 Myr. \citet{eistrup2018molecular} looked at the evolution of abundances in an evolving disc till 7 Myr and found that consideration of the disc evolution is quite necessary and complicates the case of linking exoplanetary atmospheric markers to formation locations by changing both the gas and solid C/O ratios over time significantly. Disc masses are also hypothesized to be much higher within a Myr \citep{manara2018protoplanetary, appelgren2020dust, miguel2020diverse, dash2020planet}, and hence even by the timescale used in this static model, significant disc evolution could have occurred which makes the case of a static disc assumption suspect.
\\
\\
In our work, we also do not consider the case of contamination by solids. However, such contamination can greatly change the results by changing the effective C/O ratio of the envelope. In the case of a static disc, looking at the effect of contamination from disc solids can be gauged by whether the solid C/O ratios are overall greater or less than the gas C/O ratios. For the case of molecular abundances, \citet{eistrup2016setting} show that the solid C/O ratios are always less (with an upper limit of 0.4) than the gas C/O ratios. Hence, contamination would lead to a decrease of atmospheric C/O ratios and give us a location in the disc much inner compared to where it should actually be. \citet{turrini2021tracing} used this specific case of molecular initial abundances to look at effect on C/O ratios due to enhancement from solids accreted from the disc. We use their initial C/H, O/H and N/H abundances (see Table \ref{table:1}) to redo the atmospheric chemistry calculations (using our non-inverted P-T profile) and the resultant plot is shown in Figure \ref{fig:turrini}. All the plots show H$_2$O dominated features as the resultant atmospheric C/O ratios are now closer to a value of about 0.5 which is sub-solar. The one absorption feature that prominently grows with distance is the CO$_2$ feature at 4-4.1 $\mu$m. The HCN and CO$_2$ peaks (at 13-14 $\mu$m) also start bifurcating and become more prominent with increasing distances.
\\
\\
From \citet{eistrup2016setting} we can also glean that for the case of atomic initial abundances, the solid C/O ratios at locations $>$ 2 au are larger (with an upper limit of 0.55) compared to the gas C/O ratio. Contamination here would hence have the opposite effect and give us a location much further out compared to where it should actually be. In the evolving disc scenario of \citet{eistrup2018molecular}, this situation is even more complex as gas C/O ratios can evolve and become smaller than the evolved solid C/O ratios even in the case of molecular initial abundances. While considering such cases to test out model dependent results is out of scope for this study, these can be the subject of future research work.
\\
\\
Yet another limitation is that the other mode of giant planet formation - disc fragmentation from gravitational instability is also neglected by this model. \citet{hobbs2021molecular} recently did a similar analysis for that formation mechanism and found that such a model within the CO and CO$_2$ icelines can also explain the formation mechanism of HD 209458b.
\\
\\
Other limitations of this work include assumption of an initial global C/O of 0.34 for the disc chemistry calculation and the assumption that rapid gas accretion can occur in a limited region close to the disc midplane. The validity of all these have been described in detail in \citet{notsu2020composition}. Our work is intended to serve as an extension to their work and hence inherits a lot of the same limitations of the original work.

\subsection{Accretion location from spectral observations}
Although validation of the list of species in HD 209458b's atmosphere is still not very large as discussed above, the detection of H$_2$O, CH$_4$, CO, NH$_3$ and HCN presents us an opportunity to understand at which disc location HD 209458b could have accreted most of its envelope.
\\
\\
From all the different transmission spectra we have constructed by varying initial gas elemental abundances based on disc location from \citet{notsu2020composition}, we can exclude all majority H$_2$O spectra i.e. A1/A2, E1/E2, I1/I2-K1/K2, M1/M2-O1/O2 since they show spectral fingerprints of only NH$_3$, CO and CO$_2$ other than H$_2$O. The only transmission spectra which allow for presence of all of H$_2$O, CH$_4$, CO, NH$_3$ and HCN are B1/B2-D1/D2, F1/F2-H1/H2, L1/L2 and P1/P2. Except L1/L2 (C/O of 0.41), all other cases have super-solar C/O ratios. These point to a gas accretion location outside the H$_2$O iceline and more specifically between the CO$_2$ and CO icelines for molecular initial abundance cases and between the CH$_4$ and CO icelines for the atomic initial abundance cases. For the cases of an inverted P-T profile, the spectrum as a whole is characterized by majority CH$_4$ signatures while for the case of a non-inverted P-T profile it is characterized by majority HCN signatures. In both cases, it is difficult to discriminate between the difference in chemical processing of the initial disc just based on the spectrum and would require measurement of the abundances specifically by retrieval. 
\\
\\
Our specific result of a super-solar C/O ratio and formation between the CO and CO$_2$ icelines (for a core accretion formation mechanism) is in agreement with the findings of both \citet{Giacobbe2021} and \citet{hobbs2021molecular} who found a formation location beyond the CO$_2$ iceline. \citet{Giacobbe2021} found a super solar C/O ratio close to 1 and while they found the formation location to be beyond the H$_2$O iceline, they found a likely location between the CO$_2$ and CO icelines. While \citet{Giacobbe2021} did their analysis on the assumption of thermochemical equilibrium, \citet{hobbs2021molecular}'s and our analysis is based on disequilibrium chemistry from a range of initial abundances obtained from disc chemistry simulations. The lack of a difference could be because HD 209458b is still a hot planet where thermochemical equilibrium largely dominates over disequilibrium. For cooler planets, this might no longer be the case. The other specific result corresponding to L1/2 for sub-solar C/O ratios and gas accretion between the CH$_4$ and CO icelines however is new and only exists because of an anomalous behaviour caused by significant C/H depletion in the gas accreted from the disc. However, from the discussion in the section below, it is unlikely to be a candidate but is nonetheless quite interesting.
\\
\\
\citet{Giacobbe2021}'s results also allow us to streamline possible models a bit further. The abundances of molecules obtained from their analysis for HD 209458b falls mostly beyond the values used in \citet{turrini2021tracing}, as can be found from the comparison made in \citet{hobbs2021molecular} (except for H$_2$O). Considering that their initial abundances were protosolar or larger, only B1/B2 or F1/F2 are close enough to protosolar values to fall close to this range. These models would indicate gas accretion beyond the CO$_2$ iceline, but more specifically between the CO$_2$ and CH$_4$ icelines. Both models are molecular initial abundance cases and have super-solar C/O ratios. The rest of the model cases have depleted initial elemental abundances and would require enrichment from solids. As discussed in section 4.3, enrichment using molecular initial abundance with low ionization cases were utilized in \citet{turrini2021tracing} and the resultant spectra from Figure \ref{fig:turrini} shows that HCN, CO, CO$_2$ start getting detected from T3-T6. This indicates formation beyond 19 au (i.e. beyond the CO$_2$ iceline which is at 10.5 au for \citet{turrini2021tracing}'s disc model), with enrichment of the accreted envelope by disc solids while the core is migrating. In section 4.3, we also discussed that atomic initial abundances and the case of evolving discs will complicate things even further. Looking at those cases however is out of this work's scope.

\subsection{Observability with JWST}
\begin{figure*}
    \centering
    \includegraphics[width=2\columnwidth]{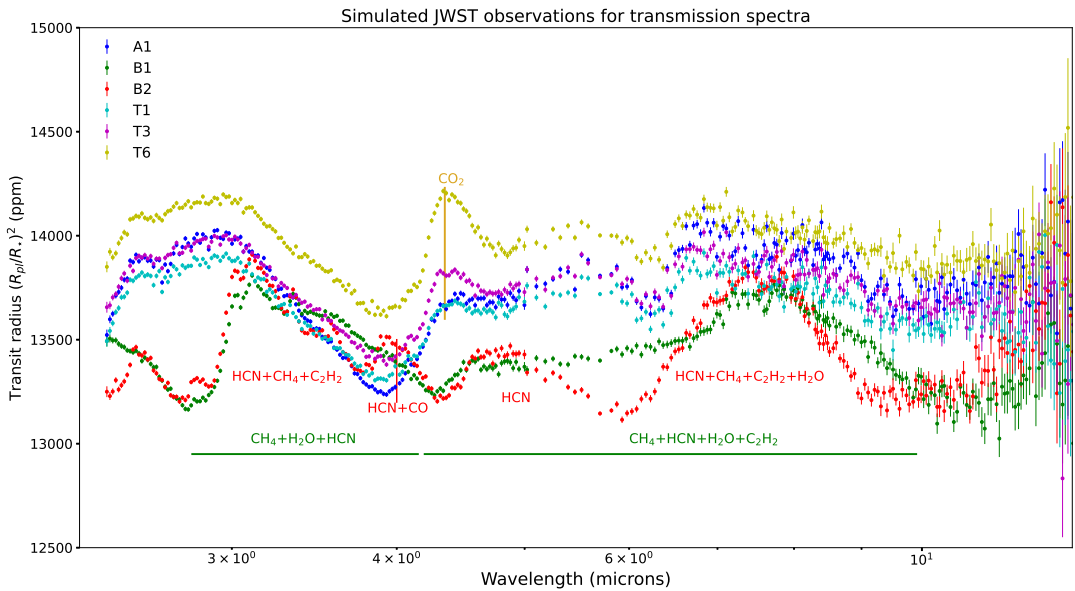}
    \caption{JWST simulated observations for transmission spectra for some selected models. A1, T1, T3 and T6 are H$_2$O dominated spectra but the increase in prominence of the CO$_2$ feature with increasing enrichment due to disc solid accretion is very clear i.e. it increases from A1 (no enrichment - only disc gas contribution) to T6 (both disc gas+solid contribution). B1 is the CH$_4$ dominated spectrum and specific molecular features are indicated in green. B2 is the HCN dominated spectrum and its associated features are indicated in red.}
    \label{fig:10}
\end{figure*}
\begin{figure*}
    \centering
    \includegraphics[width = 2\columnwidth]{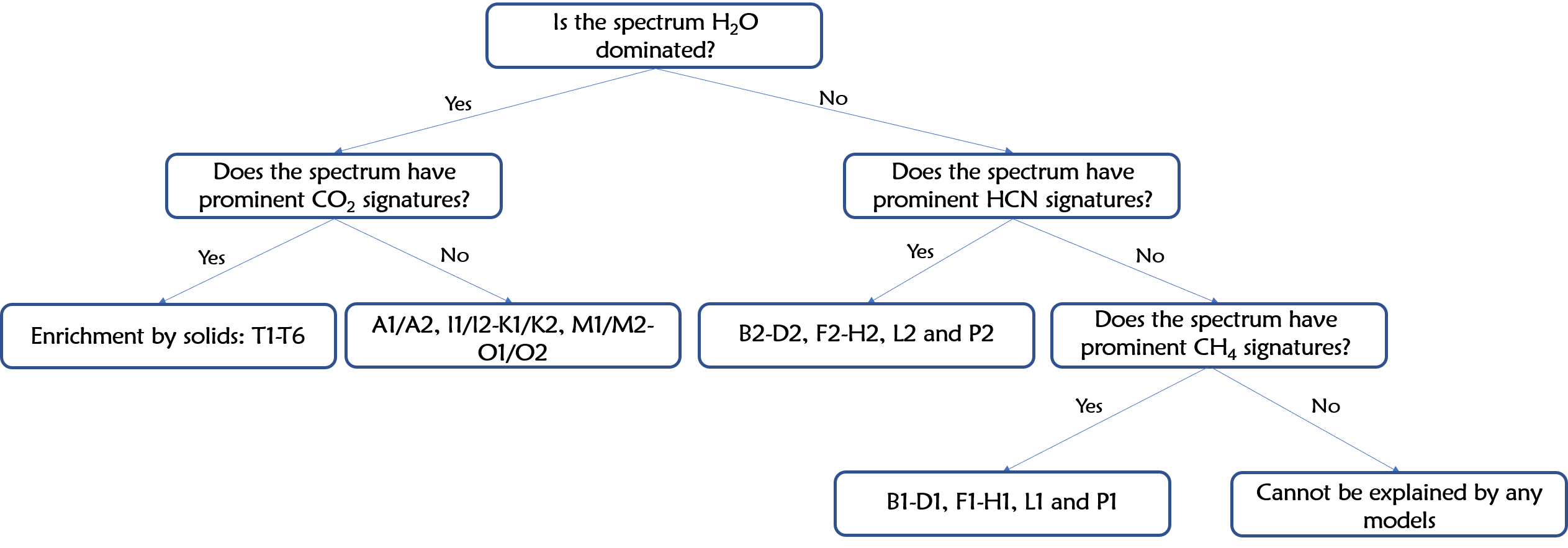}
    \caption{A flowchart showing the possible model scenarios with observational features that could be potentially inferred using JWST.}
    \label{fig:flow}
\end{figure*}

We now try to see if the three spectral cases (majority H$_2$O, majority CH$_4$ and majority HCN) that act as proxy for formation locations for this exoplanet can be differentiated using JWST observations. We have described the methodology of simulating such observations in Section 2.4. Since both B1/B2 and F1/F2 have very similar C/O ratios, initial gas phase chemical abundances and similar spectral plots, we just simulate the observation for the case of B1 and B2. We also plot T1, T3 and T6 to see if they can be differentiated as well. We simulate only the transmission spectrs because it is clear from our work that they have more prominent features.
\\
\\
From Figure \ref{fig:10}, it is easy to see that the important spectral features in our candidate spectra are indeed differentiable in transmission spectra. The CO$_2$ and CO peaks at 4-4.2 $\mu$m are very pronounced in the H$_2$O dominated spectra (A1, T1, T3 and T6). A1, T1 and T3 might be overall hard to distinguish, but T6 is distinguishable from the rest, due to the very prominent CO$_2$ feature. For the CH$_4$ dominated spectrum B1, the CH$_4$+HCN+H$_2$O+C$_2$H$_2$ broad peak from 5-9 $\mu$m is easily distinguishable from the other spectra. The narrow HCN+CO and the slightly broad HCN peaks at 4 and 5 $\mu$m respectively are also easily made out. Likewise, for the HCN dominated spectrum B2, it is easy to distinguish the dip at 6 $\mu$m and then the narrower peak due to HCN+CH$_4$+C$_2$H$_2$+H$_2$O till 9 $\mu$m. It is however slightly difficult to make out the NH$_3$ peak at 10 $\mu$m for most spectra, especially in the H$_2$O dominated spectra. Overall, it seems likely that a simple inspection of the transmission spectra can provide ways to distinguish and observationally validate the potential linkage of the atmospheric composition of HD 209458b to its gas accretion location in the disc. This is summarized in a flowchart shown in Figure \ref{fig:flow}.

\section{Conclusions}
Chemistry in protostellar discs produces radial variations in the gas phase abundances of elements including N, C and O. In this work, we look at the specific case of the hot Jupiter HD 209458b and use the radially dependent abundances predicted by the disc chemical kinetics model of \citet{eistrup2016setting} as initial abundances to predict its atmospheric chemistry. We evolve its atmospheric composition using a wrapper that couples disequilibrium chemistry with a coupled chemical kinetics (VULCAN)-radiative transfer (petitRADTRANS). We also consider the case of atmospheric evolution using two different P-T profiles: a thermally inverted profile used in \citet{moses2011disequilibrium} and a thermally non-inverted profile constructed using the self-consistent radiative transfer code HELIOS \citep{malik2017helios}.
\\
\\
We find that differences in initial abundances manifest in different mixing ratios of spectrally significant species in the atmosphere. HCN, C$_2$H$_2$, CH$_4$ and NH$_3$ are more abundant when vertical mixing, diffusion and photochemistry are enabled to use disequilibrium chemistry, compared to equilibrium chemistry mixing ratios simulated using TEA in \citet{notsu2020composition}. Sub-solar or super-solar C/O ratios (taken together with C/N ratios at times) are good indicators of relative abundances of molecules in the atmosphere if the absolute abundances of C/H and O/H are not very depleted i.e. $>\sim$ 10$^{-6}$. Depletion below this threshold can result in anomalous cases where a sub-solar C/O ratio can have prominent HCN and C$_2$H$_2$ mixing ratios. Hence, the C/O ratio when combined with C/H or O/H or the C/N ratio indicates the location where most of the gas in the exoplanet's envelope was accreted from the disc.
\\
\\
We generate synthetic transmission and emission spectra for all our atmospheric disequilibrium chemistry outputs and find that differences in the chemistry manifest clearly in different spectral signatures of H$_2$O, CH$_4$, HCN and CO in both kinds of spectra. Transmission spectra for almost all chemistry outputs are full of features and are broadly divided into three types: majority H$_2$O, majority CH$_4$ and majority HCN-CN. We then compare what species have been validated in HD 209458b's atmosphere to the spectral fingerprints of our plots and found that this requirement could possibly be satisfied by spectra from four model cases: B1/B2 and F1/F2. These correspond to cases of gas accretion between the CO$_2$ and CH$_4$ icelines for both low and high ionizations respectively for inverted and non-inverted P-T profiles. Most of the other models with similar molecular detections are too depleted in elemental abundances compared to observations from \citet{Giacobbe2021}. Our results are in agreement with the result of \citet{Giacobbe2021} as well as \citet{hobbs2021molecular}. However, we also find that contamination by disc solids can induce a massive change in the spectra by changing the C/O values of the envelope. Judging the extent of this contamination is very model dependent and can vary depending on whether one assumes molecular or atomic initial abundances. It also depends on whether the disc model is static or evolving. Hence, while we get a possible formation location using a simplified scenario, we motivate the use of such model dependent cases to judge the true extent of change due to disc solid contamination for future works.
\\
\\
Finally, we simulate observing our model atmospheres with the JWST instruments NIRCam and MIRI LRS at wavelengths 2.4-13 $\mu$m and detect the fingerprints of several species and find that we can distinguish between the three major spectra types found in our simulations i.e. majority H$_2$O, majority CH$_4$ and majority HCN(+CN). This shows that investigating hot Jupiters' formation histories through their atmospheric compositions will be viable with JWST and provide valuable insights into exoplanets' formation environments.

\software{VULCAN \citep{tsai2017vulcan, tsai2021comparative}, petitRADTRANS \citep{molliere2019petitradtrans}, HELIOS \citep{malik2017helios} and PandExo \citep{batalha2017pandexo}}

\section*{Acknowledgements}
This manuscript benefited greatly from the helpful comments of Subhanjoy Mohanty. LM thanks M. E. Ressler for discussions concerning the JWST observation. We acknowledge particularly insightful discussions on HELIOS package with Matej Malik. L.M. acknowledges the financial support of DAE and DST-SERB research grants (SRG/2021/002116 and MTR/2021/000864) of the Government of India. This research was carried out in part at the Jet Propulsion Laboratory, which is operated for NASA by the California Institute of Technology. We would like to thank the anonymous referee for constructive comments that helped to improve the manuscript.

\bibliography{example}{}
\bibliographystyle{aasjournal}

\newpage

\appendix

\begin{deluxetable*}{ccccccc}[h]
\tablewidth{0pt}
\tablecaption{Model-wise distribution of reaction pathways (type is denoted by a numerical value) at P = 1 mbar. The numerical values for the reaction pathway types are labelled in Figures \ref{fig:A1}, \ref{fig:A2}, \ref{fig:A3}, \ref{fig:A4} and \ref{fig:A5}. \label{table:At}}
\tablehead{
\colhead{Model} & \colhead{CH$_4$-C$_2$H$_2$ scheme} & \colhead{CH$_4$-CO scheme} &  \colhead{CH$_4$-HCN scheme} & \colhead{NH$_3$-HCN scheme} &  \colhead{N$_2$-NH$_3$ scheme} & \colhead{C/O}}
\startdata
A1/A2 & 1/3 & 1/1 & 1/1 & 2/2 & 1/2 & 0.35\\
B1/B2 & 2/5 & 1/2 & 1/1 & 2/2 & 1/1 & 0.81\\
C1/C2 & 2/5 & 1/2 & 1/1 & 2/2 & 1/1 & 1.27\\
D1/D2 & 2/5 & 1/2 & 1/1 & 2/2 & 1/1 & 1.00\\
E1/E2 & 1/3 & 1/1 & 1/1 & 2/2 & 1/2 & 0.35\\
F1/F2 & 2/5 & 1/2 & 1/1 & 2/2 & 1/1 & 0.82\\
G1/G2 & 2/5 & 1/2 & 1/1 & 2/2 & 1/1 & 0.94\\
H1/H2 & 2/2 & 1/3 & 1/1 & 2/2 & 1/1 & 1.00\\
I1/I2 & 1/3 & 1/1 & 1/1 & 2/2 & 1/2 & 0.35\\
J1/J2 & 1/3 & 1/1 & 1/1 & 2/2 & 1/2 & 0.35\\
K1/K2 & 1/4 & 1/1 & 1/2 & 2/1 & 1/2 & 0.13\\
L1/L2 & 2/2 & 1/1 & 1/1 & 2/2 & 1/1 & 0.41\\
M1/M2 & 1/3 & 1/1 & 1/1 & 2/2 & 1/2 & 0.35\\
N1/N2 & 1/3 & 1/1 & 1/1 & 2/2 & 1/2 & 0.35\\
O1/O2 & 1/4 & 1/1 & 1/2 & 2/1 & 1/2 & 0.25\\
P1/P2 & 2/2 & 1/2 & 1/1 & 2/2 & 1/1 & 0.89\\
\enddata
\end{deluxetable*}

\begin{figure*}
    \centering
    \includegraphics[width=1.5\columnwidth]{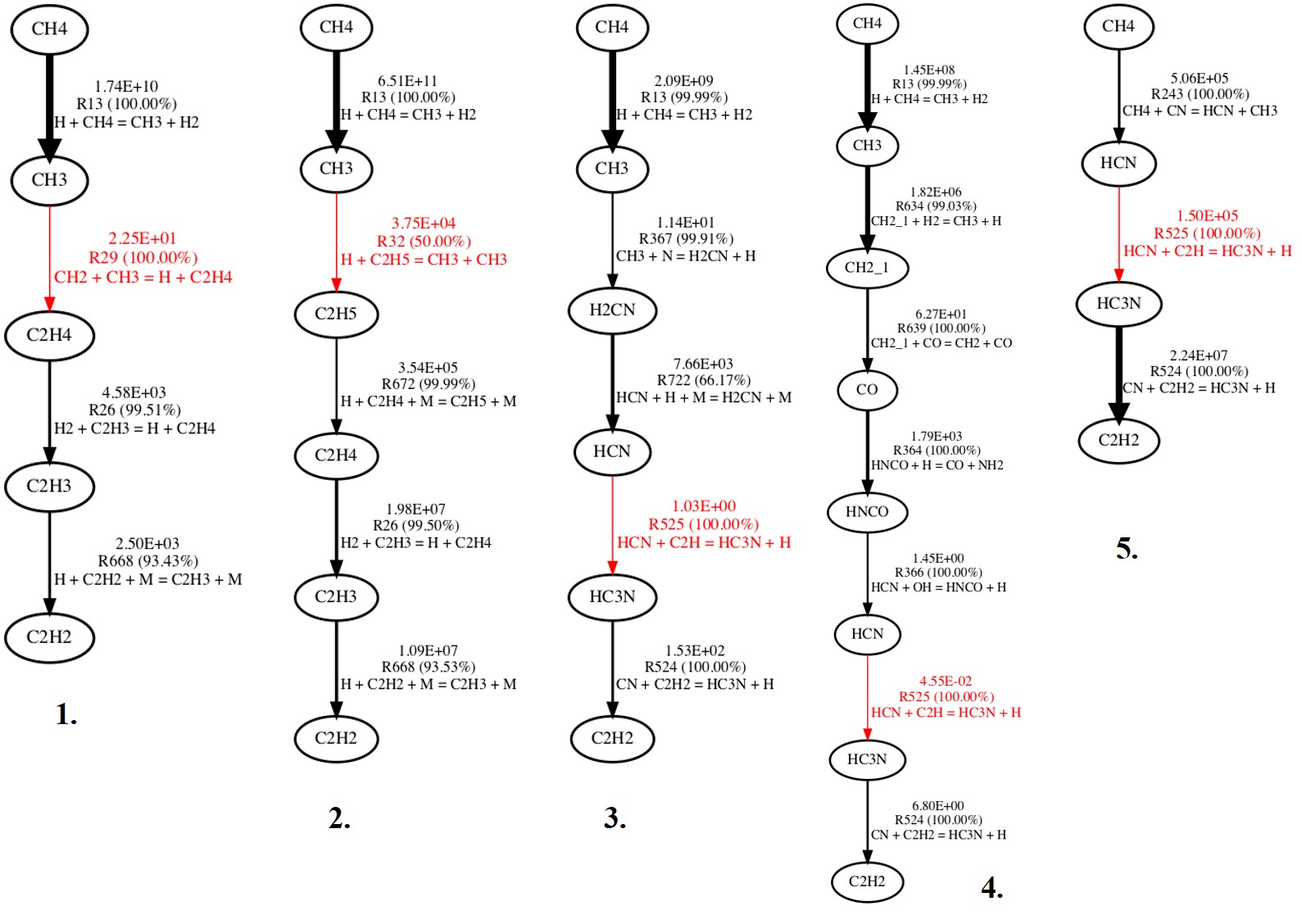}
    \caption{Examples of the types of CH$_4$-C$_2$H$_2$ reaction pathways possible in our models. Thicker lines represent faster reaction rates  which are denoted by the first-row numbers shown in cm$^{-3}$s$^{-1}$ and the red lines are the rate-limiting steps. RX denotes the reaction number from the VULCAN reaction network. An odd RX shows a forward reaction and an even RX shows the opposite. The percentage of contribution to the interconversion is also provided. The values of the rates and percentages will change from model to model while the pathway denoted by a number (below each) is representative.}
    \label{fig:A1}
\end{figure*}

\begin{figure*}
    \centering
    \includegraphics[width=1\columnwidth]{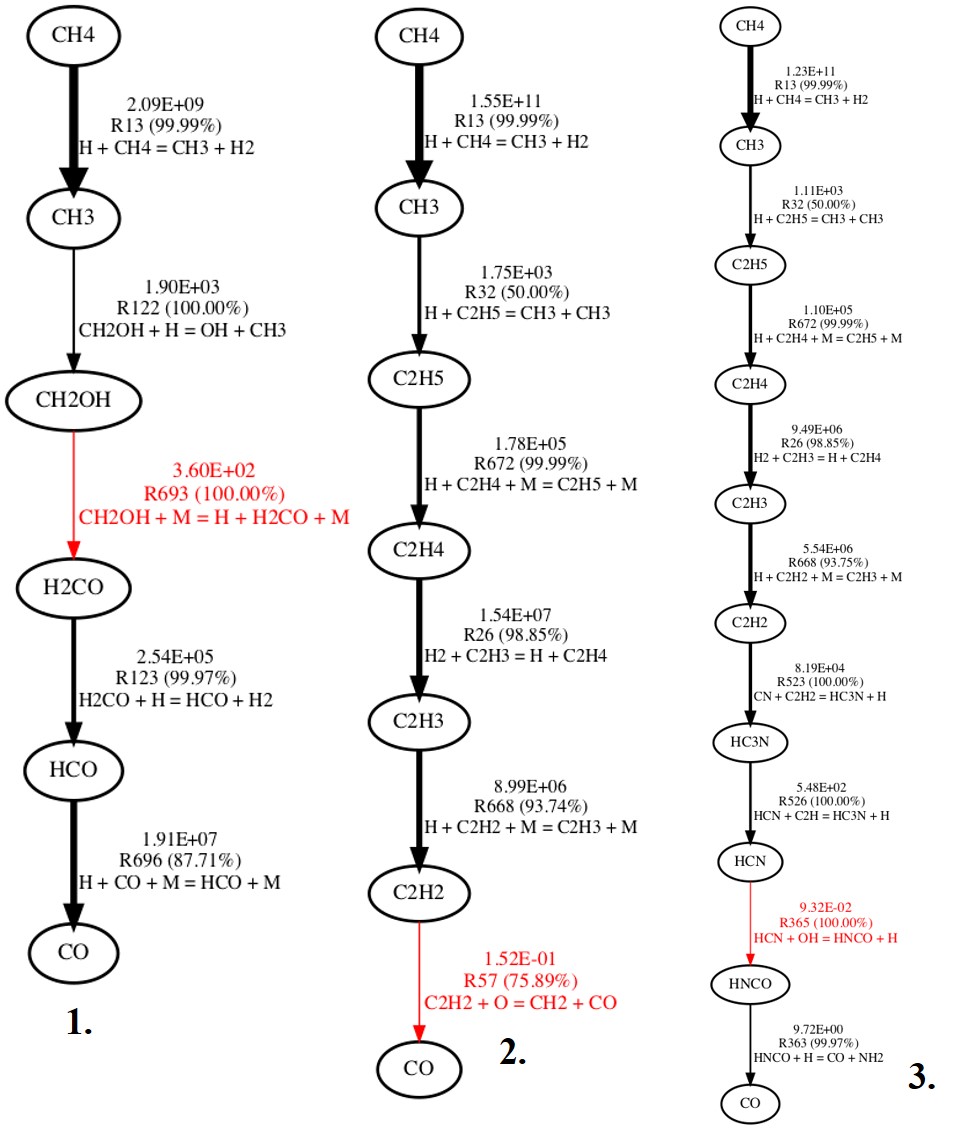}
    \caption{Same as Figure \ref{fig:A1} but for the CH$_4$-CO reaction.}
    \label{fig:A2}
\end{figure*}

\begin{figure*}
    \centering
    \includegraphics[width=0.7\columnwidth]{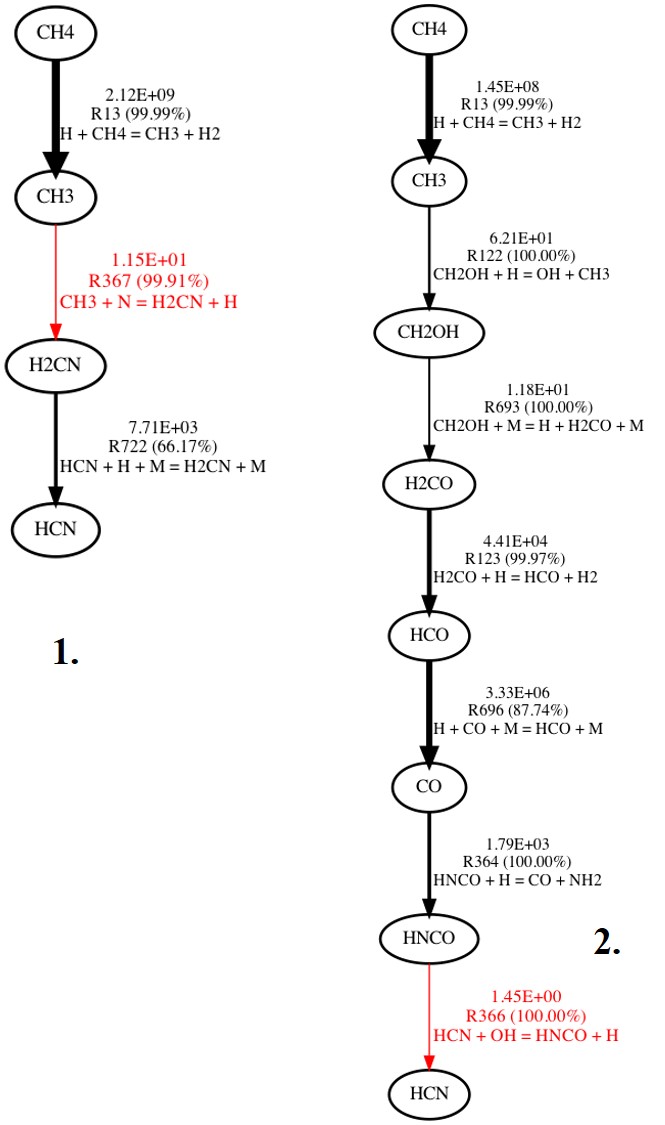}
    \caption{Same as Figure \ref{fig:A1} but for the CH$_4$-HCN reaction.}
    \label{fig:A3}
\end{figure*}

\begin{figure*}
    \centering
    \includegraphics[width=0.6\columnwidth]{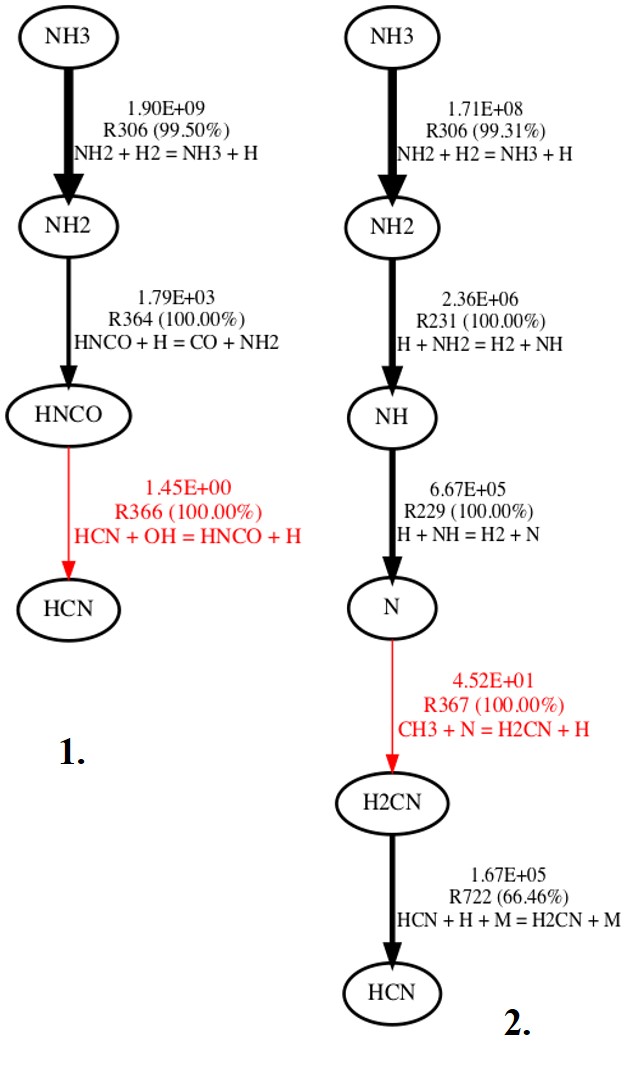}
    \caption{Same as Figure \ref{fig:A1} but for the NH$_3$-HCN reaction.}
    \label{fig:A4}
\end{figure*}

\begin{figure*}
    \centering
    \includegraphics[width=0.6\columnwidth]{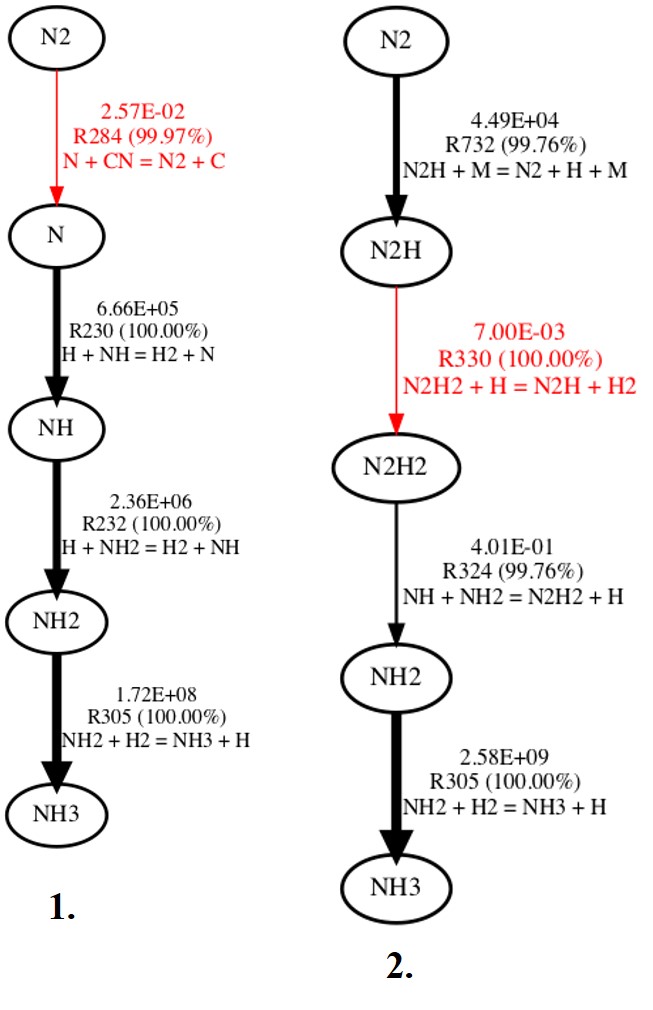}
    \caption{Same as Figure \ref{fig:A1} but for the N$_2$-NH$_3$ reaction.}
    \label{fig:A5}
\end{figure*}



\end{document}